\documentclass[12pt]{article}
\usepackage{a4}
\usepackage{cite}
\usepackage{amsfonts}
\begin{document}
\title{On Generalized Uncertainty Principle}
\author{}
\date{Bhupendra Nath Tiwari\thanks{\noindent E-mail address: bntiwari.iitk@gmail.com}\\
\vspace{0.250cm}
INFN-Laboratori Nazionali di Frascati\\
Via E. Fermi 40, 00044 Frascati, Italy.\\ \vspace{0.5cm} }
\maketitle
\begin{abstract}
We study generalized uncertainty principle through the basic
concepts of limit and Fourier transformation and analyze both the
quantum theory of gravity and string theory from the perspective
of complex function theory. Motivated from the noncommutative
nature of string theory, we have proposed a UV/IR mixing dependent
function $ \tilde{\delta}(\Delta x,\Delta k, \epsilon) $. 
For a given $ \tilde{\delta}(\Delta x,\Delta k, \epsilon) $, 
we arrived at the string uncertainty principle from the analyticity condition 
of a complex function, which depends upon UV cut-off of the theory.
This non trivially modifies the quantum measurements, black hole
physics and short distance geometries. The present analysis is based
on the postulate that the Planck scale is the minimal length scale
in nature. Furthermore, our consideration is in perfect agreement
with the existence of the maximum length scale in nature. 
Both of the above length scales rely only upon the analysis of  
$ \tilde{\delta}(\Delta x,\Delta k, \epsilon) $ and do not directly 
make use of any specific structure of the theory or Hamiltonian. 
The Regge behavior of the string spectrum and the quantization of 
the horizon area of a black hole are natural consequences of the 
function $ \tilde{\delta}(\Delta x,\Delta k, \epsilon) $. It is hereby
anticipated that $ \tilde{\delta}(\Delta x,\Delta k, \epsilon) $ 
contains all possible corrections operating in nature, and 
thus a promising possibility to reveal important clues towards 
the geometric origin of $M$-theory. \\

\vspace{0.250cm}


\vspace{0.250cm}


\end{abstract}

\newpage
\section{Introduction}

The analysis of the uncertainty principle has led to several imminent 
insights into the classical and quantum aspects of the gravity theories
\cite{GreenSchwarzWitten},\cite{Polchinski},\cite{AlainConnes}. 
As a possible candidate of the quantum gravity, the string theory
has offered a number of exciting developments in physics and mathematics.
For example, from the perspective of general relativity, field theories and short distance geometries,
there have been interesting developments in the string theory, \textit{e.g.},
$D$-brane physics \cite{Polchinski}. In this sense, the string theory provides
the best clue towards the grand unification of the fundamental forces of nature \cite{GreenSchwarzWitten},
\textit{viz.} the question \textit{``how to obtain a unified theory that describes
all the laws of nature"}. Further, the string theory has led many new discoveries
both in the fundamentals of theoretical physics and pure mathematics
\cite{GreenSchwarzWitten},\cite{Polchinski},\cite{AlainConnes}. 
However, the tools needed to fully understand the string theory configurations
are yet in the path of their developments. Thus, the question of the ultimate fundamental 
building blocks and the ultimate physical reality of nature yet remains open.
It is not known in what direction there would be its final solution,
or even whether such a final objective answer can be at all expected. 

In this paper, we provide the most general form of the uncertainty inequality. 
As per the one of the main thrust of the present work, we show that such a 
consideration offers an important clue towards the resolution of ultraviolet (UV) 
and infrared (IR) mixing problem and thus the possible quantum gravity candidate
theories. In this perspective, we find that the short distance properties 
of physical spacetime manifold can be expressed in terms of an energy 
dependent function of the coordinates and momenta. The first order
approximation of the present consideration leads to the noncommutative
behavior of string theory. In a generic sense, our analysis applies 
to all possible quantum theories. Furthermore, we demonstrate that the 
Regge behavior of the stringy spectrum, black hole horizon area 
quantization, quantum behavior of the $(3+1)$ spacetime, scaling properties,
Fourier transformation, Wigner transformation, distribution theories 
and discreate manifolds arise as a natural consequence of the generalized 
uncertainty principle. From such an unified consideration, it appears that 
all the corrections operating in the finally unified field (string) theory may be revealed 
from the specific examinations of $ \tilde{\delta}(\Delta x,\Delta k, \epsilon) $.  
In section $3$, we shall give an explicit consideration towards the proposition of
the function $ \tilde{\delta}(\Delta x,\Delta k, \epsilon) $. In this sense, 
we shall illustrate in section $4$ that $ \tilde{\delta}(\Delta x,\Delta k, \epsilon) $ 
offers an important clue towards a geometric origin of the fundamental physical theories. 
Thus, in order to understand the limiting field theory configurations,
we explore the complex analysis framework for the quantum nature of physics 
under the higher order corrections, \textit{e.g.}, generalized uncertainty 
principle corrections and string theory corrections.

A priori, we know that the short distance physics is not very well understood.
Thus, in order to adequately describe small scale structures of the spacetime,
we need to modify the standard classical continuum geometries by their quantum counterparts.
Such an example follows from the Connes non-commutative geometry \cite{AlainConnes}. 
In other words, the small scale physics indicates the fact that an extension 
of quantum mechanics might be required in order to accommodate the gravity.
Moreover, the existence of the duality symmetries in non-perturbative string theory \cite{EdwardWitten96},\cite{EdwardWitten97}
shows that the strings do not distinguish between the small and the large spacetime scales.
Beyond the Planck scale energies, such a consideration requires a modification of
the standard Heisenberg uncertainty principle in order to incorporate the fact that 
the size of the underlying strings grows with the momenta instead of falling off. 
As a result of the string theory, a consideration of such a $T$-dual description of  
spacetime has been introduced by Witten, see \cite{EdwardWitten96},\cite{EdwardWitten97}
and references therein. In this consideration, the very concept of the spacetime changes 
its meaning beyond the Planck length scale. Subsequently, the standard Heisenberg
uncertainty principle calls up for its modification. Thus, the main thrust of the
present work is to explain the notion of the completely generalized uncertainty
inequality. From the viewpoint of complex function theory, such an analysis involves 
the fundamental string length $ l_s $, at  the first order. As the function of $l_s$, 
we show in section $3$ that the first order faithful interpolation of the function 
$ \tilde{\delta}(\Delta x,\Delta k, \epsilon) $ yiels to the following form of the 
uncertainty relation $ \Delta x^{\mu} \geq \frac{\hbar}{\Delta p^{\mu}} 
+ \frac{l_s^2}{\hbar} \Delta p^{\mu} $, where the speed of light $c$ is 
henceforth taken to be $ c= 1 $.

Indeed, Castro \cite{CarlosCastro} has conjectured that the special theory of scale relativity, 
as recently proposed by Nottale \cite{LNottaleIJMP1},\cite{LNottaleIJMP2}, 
must play a fundamental role in the string theory. Specially, it demonstrates 
that there is the universal, absolute and impassable length scale in nature, 
which remains invariant under the dilatations. The corresponding lower limit
of such a scale is expected to be the order of Planck length scale. 
In the viewpoint of the string duality principles \cite{EdwardWitten96},\cite{EdwardWitten97}, 
the fundamental scales of nature are determined by constraints, which are fixed at both
the small and large scales. Applying the scale relativity principle to the universe,
one arrives at the proposition that there must exist an absolute, impassable, upper
length scale in nature. Such a length scale $L$ must remain invariant under the dilatations.
Particularly, the expansion of the universe indicates that the scale $L$ inherits all possible
physical properties at the infinity. This upper scale $ L $ defines the radius of the universe. 
Thus, the length $L$ becomes invariant under the dilations, when it is seen at its own resolution scale.


Recently, it follows that the quantum gravity measurements are
governed by the generalized uncertainty principles. Evidences from the
string theory \cite{GreenSchwarzWitten},\cite{Polchinski}, quantum geometry 
\cite{AlainConnes},\cite{LNottaleIJMP1}, and black hole physics
\cite{PChenandRJAdler},\cite{RJAdlerPChenandDISantiago} suggest
that the standard Heisenberg uncertainty principle needs certain
modification(s). These evidences have an origin on the quantum
fluctuations of the background metric tensor. Thus, the generalized
uncertainty principle provides an existence of the minimal length
scale in nature. At the first order approximation, we show that
such a minimum scale is of the order of Planck length scale.
From the perspective of a generalized uncertainty principle,
Adler et. al. \cite{PChenandRJAdler},\cite{RJAdlerPChenandDISantiago}
have considered the issue of black hole remnants. In this framework, 
they have shown that the generalized uncertainty principle prevents
total evaporation of small black holes. Whilst, the Bekenstein-Hawking
approach \cite{BH1},\cite{BH2},\cite{BH3} allows the total evaporation of 
micro black hole configurations. In this perspective, the generalize 
uncertainty principle indicates the nature of the quantum gravitational 
corrections to the black hole thermodynamics.


Due to the uncertainty inequality, the coordinates and the corresponding 
canonical momenta cannot both be simultaneously specified after the quantization.
Moreover, the associated phase space as the total physical state space must be
modified for the non-zero Planck constant. For a given quantum mechanical configuration, 
the Hilbert space of wave functions suggests that the space of states can be 
characterized by the principle of superposition of the corresponding quantum states. 
Similarly, from the viewpoint of the validity of spacetime uncertainty relation, 
we expect that certain modifications exist into the notion of spacetime geometry. 
Along this line of thought, the string theory can be taken as some sort of `quantum geometry'.
However, at present, it is difficult to formulate this kind of ideas concretely. 
Before the modern duality perspective of the string theory \cite{EdwardWitten96},\cite{EdwardWitten97}, 
there have been several attempts to generalize the local field theories. Based on similar ideas,
it has been difficult to include the gravity in an appropriate framework of the
quantum gauge theories. For example, let the spacetime coordinates be
operators instead of ordinary $C$ numbers, then both the above coordinates and
the associated momenta can be treated as operators acting on certain Hilbert space. 
In the viewpoint of quantum mechanics, such an idea has been recently seen
in the context of string theory. With an appropriate assumption of background fields,
it arises as an akin limit of the `noncommutative field theory' 
\cite{DouglasNekrasov},\cite{Yoneya},\cite{SeibergWitten}. However, these limits 
neglect the crucial extendedness of the strings along the longitudinal directions. 
At this juncture, we do not have notably new insight on the spacetime uncertainties 
characterized by the string length parameter $ l_s $. In this concern, the microscopic 
perspective of noncommutative field theories and short distance geometries opens new
research avenues for further investigations.


As a matter of fact, the Heisenberg uncertainty principle needs to
be reformulated due to the noncommutative nature of spacetime manifold,
at the Plank length scale. Consequently, in the limit of extreme quantum gravity,
there exists a minimal observable distance of the order of Planck length scale, 
where all the short distance measurements are governed. In the context of string theory, 
such an observable minimal distance can be estimated from a generalized uncertainty 
principle $ \Delta x \geq \frac{\hbar}{\Delta p} + $ constant.\ $ G \Delta p $. 
Thus, the existence of the minimal length scale in nature 
\cite{KouroshNozariandSHamidMehdipour},\cite{LGaray},\cite{AKempfandGMangano},\cite{CameliaLukierskiNowicki}
brings out the fact that the Planck length $l_p$ satisfies the following 
inequality $ \Delta x \geq \frac{\hbar}{\Delta p} + \frac{\alpha^{\prime}
l_p^2}{\hbar} \Delta p $. This is because the minimal length scale exists
in nature, and so there are possibilities to correct the standard
Heisenberg uncertainty principle, as a limiting characterization of
the generalized uncertainty principle. In the string theory set up,
the higher order contributions are non zero in any general theory,
however the minimal length is governed by the stringy parameter
$ \alpha^{\prime} $. For a momentum dispersion $ \Delta p $, 
such a generalized uncertainty principle shows that the minimum
coordinate dispersion $ \Delta x $ is at least non zero, 
as long as the first two terms on the right hand side of the
above uncertainty relation are non zero. It is worth mentioning that 
the generalized uncertainty principle, \textit{i.e.}, an existence of 
the minimum length scale in nature, is further motivated from the study 
of the short distance behavior of (i) string theory 
\cite{GVenezianoEurophys},\cite{DGrossPFMende},\cite{DJGrossandPFMende},\cite{DAmatiMCiafaloniandGVeneziano},\cite{KKonishiGPautiPProvero},\cite{DAmatiMCiafaloniandGVeneziano1},
(ii) black hole physics \cite{MMaggiore} and (iii) de Sitter spaces \cite{HSSnyder}.


From the perspective of $D$-brane physics, the spacetime uncertainty principle 
gives an intriguing qualitative characterization of nonlocal and noncommutative
nature of the short distance spacetime structures in string theory. 
For example, Yoneya \cite{TamiakiYoneya} has analyzed space-time uncertainty 
relations and possible approaches to $D$-brane field theory.
Recently, there have been several investigations which take an account
of the generalized uncertainty principle approach into the physics of
$D$-brane field theories by emphasizing the key issues lying in the 
background space-time manifold, see \cite{TYoneya} and references therein.
Further motivations arise from the quantum mechanical propagator of a bosonic 
$p$-brane configuration obtained in the quenched minisuperspace approximation \cite{AntonioAurilia}.
This suggests an amusing possibility of the novel and unified description
of $p$-brane dynamics with different dimensionalities \cite{AntonioAurilia}.
Hereby, the background metric emerges as the quadratic form on a Clifford manifold.
With an account of the Clifford line element, the substitution of standard Lorentzian
spacetime metric changes its very structure as per the spacetime fabric. 
This follows from the fact that the new metric is built out of the minimum length,
below which it is impossible to resolve the distance between the chosen two points.
Furthermore, it is worth mentioning that the introduction of the Clifford line element 
extends the usual relativity of motion to the case of relative dimensionalities.
Near the Plank length scale, such a consideration involves all possible values 
of $p$, and thus a collection of $p$-branes contributes significantly towards
the understanding of short distance structures of the physical $(3+1)$ spacetime.


The string theory corrections to the original Heisenberg uncertainty
principle follow further from the quantum mechanical wave equations on
a noncommutative Clifford manifold, where all dimensions and signatures
of the spacetime are taken on the same footing \cite{AntonioAurilia},\cite{CastroCarlos}. 
In this concern, Castro \cite{CastroCarlos} has outlined a new relativity principle 
towards the fully covariant formulation of $p$-brane quantum mechanical loop wave
equations, where the stringy uncertainty relation arises naturally. 
In fact, there exists one to one correspondence between the nested hierarchy
of $p$-loop histories encoded in terms of the underlying hypermatrices and the 
associated wave equations written in terms of the Clifford algebra valued multivector 
quantities. This allows one to write the quantum mechanical wave equations
associated with a hierarchy of nested $p$-loop histories embedded
in the chosen $D$-dimensional target spacetime with a single quantum
mechanical functional wave equation. The corresponding line elements
of the $p$-brane histories lies in a noncommutative Clifford manifold 
of $ 2^D $ dimension having $ p= 0, 1, 2, 3 \ldots D- 1 $,
where $ D-1 $ is the maximum value of $p$ that saturates the
dimension of the embedding spacetime manifold \cite{Castro}.
In such a C-space, the components of a co-ordinate $ x $ and the 
corresponding conjugate momentum $ p $ must not be interpreted 
as the ordinary vectors of spacetime, but they need to be treated
as the Clifford algebra valued multivectors. Near the Planck
length scales, such multivectors ``coordinatize" the underlying 
noncommutative Clifford manifold. In fact, the noncommutativity 
may be encoded in an effective Planck's constant $ \hbar_{eff} $,
which modifies the operators of the Heisenberg Weyl commutation algebra,
for a given $\{ \widehat{x}, \widehat{p} \}$. Consequently, while keeping 
the first two terms in the expansion of $ \hbar_{eff}(k) $, Ref \cite{Castro} 
shows that such a consideration obeys the standard string uncertainty relation
$ \Delta x \geq \frac{\hbar}{\Delta p}+ \frac{\beta_1 l_p^2}{4 \hbar} (\Delta p)$.
This follows from the inherent noncommutative nature of the underlying Clifford manifold,
which could reshuffle a chosen loop history into the other loop histories,
membrane histories, a chosen membrane history into a $p$-brane history.
More generally, such a consideration can transform a $p$-brane history
into suitable combinations of all possible $p$-brane histories and thus defines
the building blocks of the theory. This idea takes one point particle to $p$-branes,
in which each brane is made out of all possible other $p$-branes.
In this case, the Lorentz transformations in a C-space involve hypermatrix
changes of the coordinates in $p$-brane quantum mechanics \cite{CastroCarlos}.


Thus, as per the above considerations, we have made an effort to bridge up both 
the string theory and the quantum mechanics within an extended framework of the 
Heisenberg uncertainty principle. We consistently define the most general 
uncertainty principle expression than those existing in the literature 
\cite{Yoneya},\cite{KouroshNozariandSHamidMehdipour},\cite{LGaray},\cite{AKempfandGMangano},\cite{CameliaLukierskiNowicki},\cite{DGrossPFMende},\cite{DJGrossandPFMende},\cite{KKonishiGPautiPProvero},\cite{DAmatiMCiafaloniandGVeneziano},\cite{TamiakiYoneya},\cite{TYoneya},\cite{AntonioAurilia},\cite{Castro},\cite{DAmatiMCiafaloniandGVeneziano1},\cite{MFabrichesiGVeneziano},\cite{RGuidaKKonishiPProvero},\cite{YoneyaTamiaki},\cite{BangBerger},\cite{AntalJevickiTamiakiYoneya}.
From the complex analysis perspective, we notice why the upper scale of nature must 
appear in the fundamental equation? The importance of this analysis is in accordance 
with the existence of noncontinuous maps in the string theory. Such notions have
been discussed by Borde and Lizzi \cite{JBordeandFLizzi}. 
In order to reproduce the results of the dual models \cite{GVeneziano},
the space of string configurations require that the string theory possesses
both the continuous and the noncontinuous square integrable maps. 
Subsequently, the size and the shape of the concerned strings have been investigated
in their ground states, in the light cone gauge theories \cite{MKarlinerIKlebanovLSusskind}.
At this point, Susskind et. al. \cite{MKarlinerIKlebanovLSusskind} have shown 
that the extrinsic curvature diverges in the case of two dimensions.
Thus, an appropriate regularization scheme is needed where the string is kept
continuous. As a matter of fact, the strings become smoother and have a divergent average size,
as the dimensionality of the spacetime increases. Such an effect is unphysical, 
since the size of strings cannot exceed the size of the universe. 
Because of this reason, the upper length scale of the universe must also appear
in the fundamental equation of nature. In fact, the four dimensional average 
curvature diverges due to kinks and cusps solutions of the constituent strings.
Thus, it is important to study the above subject matter further and analyze the
properties of correcting functions, in the light of the generalized uncertainty principle.
Having presented the reasons why the uncertainty principle is an important issue, 
thus, we explain the relevance of the stringy uncertainty principle from the
complex function theory. Mathematically, we show that the generalized uncertainty 
inequalities reveal a variety of issues for further research in the subject matter.


In the present paper, we study the effects of higher derivative stringy
and quantum gravity corrections to the limiting generalized uncertainty principles.
In particular, for a given complex function, the analysis of the holomorphic
(anti-holomorphic) function gives a simple explanation of the stringy
uncertainty principle. This introduces the problem and its motivation. 
Further, the organization of the present work has been offered in
the five sections. In section $ 2 $, we review standard uncertainty 
principle in finite dimensional Euclidean quantum mechanics. By
taking an account of the shape and size, we have shed light on the 
Heisenberg uncertainty principle for arbitrary finite set of $ L^2 $ functions. 
In section $ 3 $, motivated from the string theory and its emphasis on
the concept of limit and Fourier transformation of a complex function, 
we propose a function $ \tilde{\delta}(\Delta x,\Delta k, \epsilon) $, 
giving a resolution criteria for the UV/IR mixing problem. Furthermore,
we have perturbatively proved the proposition and outline specific
generalizations for the case of arbitrary manifold. In section $ 4 $, 
we have offered physical and mathematical motivations behind the complete
generalization of uncertainty inequalities. Such a consideration
renders to the string uncertainty principle as the first order approximation 
and its all order perturbative corrections offer a resolution criterion towards 
the UV/IR mixing problem, physics of quantum gravity, black hole physics, 
existence of the minimal and maximal length scales, role of short distance 
geometries in string theory, Fourier transformation, distribution theory
and possible discretization of the spacetime. Moreover, our completely 
generalized uncertainty principle reveals a large number of known physical 
and mathematical issues at a single stack. In this concern, we show that 
the newly proposed function $ \tilde{\delta}(\Delta x,\Delta k, \epsilon) $ 
reveals a geometric origin towards the understabding of fundamental $M$-theory. 
Finally, the section $ 5 $ contains a set of concluding issues and
remarks for a future research investigation.

\section{Measurements and Quantum Mechanics}

In this section, we review basic issues concerning the
physics and mathematics of the uncertainty principle. Namely,
we wish to begin our analysis by considering a set of needful 
fundamentals from the quantum mechanics and the measurement theory
\cite{Dirac}. To do so, we first focus our attention on the
role of a physical observable, whose string theory extension 
is offered in the next section.
On the basis of the dual nature of matter \cite{Dirac}, it is well known 
that it is macroscopically possible to exactly measure the position 
and momentum of a moving particle at a given instant. However, 
it is microscopically not possible to exactly (or with certainty)
measure both the position and momentum of a particle. A simultaneous
measurement of the position and momentum of a particle requires
the laws of wave mechanics. According to the quantum mechanical theory \cite{Dirac},
the behavior of a moving particle is defined by a wave packet, where
the particle moves with the following group velocity $ v_g = \frac{d \omega }{dk} $.
As per the Max-Born criterion \cite{Dirac}, the particle can be found anywhere within 
the wave packet. In this accordance, the position of a particle is uncertain within the 
limits of the wave packet. Furthermore, this consideration leads to the fact that the 
wave packet advances with the above group velocity $v_g$, and thus there exists 
uncertainty in the velocity or the momentum of the chosen particle.

At a given point in the configuration space, it is well known that 
the square of the amplitude of the wave packet of a particle represents 
the probability of finding the particle at the chosen point. 
For example, the periodic wave function with a constant wave
length has no uncertainty in the momentum. However, since the 
amplitude of the corresponding waves remains the same everywhere,
thus the associated probability remains the same. This shows that
the particle can exist anywhere in the wave packet, \textit{viz.}
there is infinite uncertainty in the position measurement of the
particle. Mathematically, for a given Fourier transform pair, 
the uncertainty principle follows directly from the scaling
property of the Fourier series expansion. Physically, the above
scaling can be stated as ``Time Duration $ \times $ Frequency Bandwidth 
$ \geq C $", where $ C $ is a constant being determined from the 
precision definitions of ``duration" in the time domain and 
``bandwidth" in the frequency domain. From the definition of
the Fourier transform and inverse Fourier transform, it follows
that, if the duration and the corresponding bandwidth are defined
as a pair of nonzero intervals, then one obtains that $ C=\infty $.
This is not very informative in the physical situations and thus 
motivates us for the further analysis.

In order to illustrate the above definition, let us consider the formation
of a wave packet, which is formed by the superposition of two progressive
waves of infinitesimall different angular frequencies $ \omega , \omega + d
\omega $ and propagation constants $ k, k + dk $ satisfying $ k =
\frac{2 \pi }{\lambda} $, where $ \lambda $ is the wave length. Let
the equations of the above waves be $ \varphi_1 = A sin(\omega t - k x
), \varphi_2 = A sin((\omega + \Delta \omega) t - (k + \Delta k) x
) $. Then, the theory of superposition implies that the wave packet thus
formed can be expressed as $ \varphi = \varphi_1 + \varphi_2 = 
2 A sin(( \omega+ \frac{\Delta \omega}{2})t - (k + \frac{\Delta k }{2})x ) \cdot
cos((\frac{\Delta \omega}{2})t - (\frac{\Delta k}{2}) x ) $. Or $
\varphi = 2 A cos((\frac{\Delta \omega}{2})t - (\frac{\Delta
k}{2}) x ) \cdot sin(\omega t - k x ) $. Thus, the amplitude of the
resultant wave is $ A_{res}= 2 A cos((\frac{\Delta \omega}{2})t -
(\frac{\Delta k}{2}) x ) $ and the spread in the wave packet is
equal to the half of the wave length $ \lambda_m $ of the resultant
wave. Consequently, the uncertainty in the position of the particle
is $ \Delta x = \frac{ \lambda_{m}}{2} $. As the propagation constant of
the wave packet is $ k_m = \frac{\Delta k}{2} $, and thus the definition
$ k_m = \frac{2 \pi}{\lambda_m} $ implies that $ \Delta x \Delta k = 2 \pi $.
For the purpose of subsequent discussion, let $ k = \frac{2 \pi}{\lambda} $
and $ \lambda = \frac{h}{p}$. Hence, we have $ k = \frac{2 \pi p}{h} $ 
and $ \Delta k = \frac{2 \pi \Delta p}{h} \Rightarrow \Delta x \Delta p = h $.
This is the well celebrated Heisenberg uncertainty principle in quantum
mechanics. Let us now take a more closer look at the above examination. 
From the theory of statistical deviation, an explanation arises 
from the fact that both the size and the shape of a wave packet
are important to take into the consideration. This amounts to replace the
above equality with the following inequality $ \Delta x \Delta p \geq
\frac{ \hbar}{2} $. For a particle moving in $ R^1 $, let us now 
consider the Heisenberg uncertainty principle in a rigorous 
setting of the function theory and quantum mechanics.

The usual uncertainty measurements of a one dimensional quantum
mechanical particle can be described by calculating the uncertainties 
in the position and the corresponding momentum of the particle. 
As mentioned earlier, the standard statistical deviation method leads 
to a rigorous illustration of the Heisenberg uncertainty principle. 
To do so, let $ \psi(x) $ be the normalized wave function of a particle in $ R^1 $,
then the probability of finding the particle between the position 
$ x $ and $ x+ dx $ is defined by $ \psi^{\star}(x) \psi(x) dx $. 
For the normalized  $ \psi(x) $, the expectation value of the position
$ x $ is defined by $< x >:= \int \psi^{\star}(x) x \psi(x) dx $. Then,
the uncertainty in position of the particle in $ x $ direction is given by $
\Delta x := [< \lbrace x - <x> \rbrace^2 >]^{1/2} $. Similarly, it follows
that the uncertainty in the momentum is given by $ \Delta p := [< \lbrace p -
<p> \rbrace^2 >]^{1/2} $, where $ <p> := \int \psi^{\star}(x) (-i
\hbar \frac{\partial}{\partial x})\psi(x) dx $. Without loss of the
generality, we can choose the coordinate system of our interest such that
$ < x >= 0 $ and the average momentum is also zero, \textit{i.e.}, 
we can assume that the normalized wave packet is initially centered 
at $ < x > = 0 $  and $ < p > = 0 $. Thus, we have $ (\Delta x)^2 = <x^2> $
and $ (\Delta p)^2 = <p^2> $.

Before proceeding further, let us now analyze the integral $ i
\hbar \int_{-\infty}^{+\infty}\psi^{\star}(x) \frac{d}{dx}(x
\psi(x)) $. Integrating by parts from $-\infty $ to $ +\infty
$ and applying the boundary conditions $ \psi^{\star} \psi \vert_{x= \pm
\infty}= 0 $, it turns out that we have $ i \hbar \int \frac{d \psi^{\star}}{dx} x
\psi dx = - i \hbar \int \psi^{\star}\frac{d}{dx} (x \psi(x)) dx =
- i \hbar [\int \psi^{\star} x \frac{d \psi }{dx} dx + \int
\psi^{\star} \psi .1.dx ]$. Invoking the fact that $ \int \psi^{\star} \psi dx = 1
$ for a normalized $ \psi(x) $, we arrive at the identity $ i \hbar \int \frac{d
\psi^{\star}}{dx}. x \psi dx + i \hbar \int \psi^{\star} \frac{d
\psi }{dx}x dx = -i \hbar $. Since the quantity in the right hand
side is purely imaginary, and thus the left hand side must also be 
equal to a imaginary quantity. Specifically, we find that the imaginary
part of left hand side is $ (2i \int i \hbar \frac{d \psi^{\star}}{d x}x \psi dx) = - i \hbar
$. Now taking the modulus and square of the both hand sides, we get
$ 4 \vert Im(\int i \hbar \frac{d \psi^{\star}}{d x} x \psi dx) \vert^2=
\hbar^2 $. Since $ \vert \overline{z} \vert \geq \vert y \vert$
for any complex number $ z= x+ iy $. Thus, $ \vert \int i \hbar
\frac{d \psi^{\star}}{d x} x \psi dx) \vert \geq \vert Im(\int i
\hbar \frac{d \psi^{\star}}{dx}x \psi dx) \vert $. This implies that
$ 4 \vert \int i \hbar \frac{d \psi^{\star}}{dx}x \psi dx
\vert^2 \geq \hbar^2 $. For any two complex functions $ f $
and $ g $, the Schwarz inequality $ \vert \int f^{\star} g dx \vert^2 
\leq (\int f^{\star} f dx )(\int g^{\star} g dx) $ leads to the fact that 
we have $ \vert \int i \hbar \frac{d \psi^{\star}}{dx}x \psi dx \vert^2 \leq (\int i
\hbar \frac{d\psi^{\star}}{dx}(-i \hbar \frac{d \psi}{dx}) dx
)(\int x \psi x \psi^{\star} dx) $. As per the present case, we see that the
above equation reads $ \int i \hbar \frac{d \psi^{\star}}{dx} (-i \hbar \frac{d
\psi}{dx}) dx \int  \psi^{\star} x^2 \psi dx \geq
\frac{\hbar^2}{4} $. From the definition $ <x^2>= \int
\psi^{\star} x^2 \psi dx $ and $ <p^2>= \int \vert -i \hbar
\frac{d \psi}{dx} \vert^2 dx $, one obtains $(\Delta x)^2
(\Delta p)^2 \geq \frac{\hbar^2}{4} $. In particular, we have
$ (\Delta x)  \geq \frac{\hbar}{2 (\Delta p)}$. In ordinary
one dimensional quantum mechanics, the above inequality is known
as the Heisenberg uncertainty principle.

At the short distance regimes, the notion of the
continuum spacetime is drastically modified. From the perspective of 
the string theory, such a consideration may ultimately unify general
relativity with the quantum gauge theories. In this research, we aim to
discuss possible significances of the short distance aspect of string theory. 
From the perspectives of the uncertainty principle and complex analysis, 
one of the distinguishing feature of any quantum theory,
when it is compared with the corresponding classical physics, 
is that there exists nonzero quantum fluctuations. In classical physics, 
a physical state can in principle be exactly determined with sufficient 
knowledge of the state. Thus, at a given time, one can exactly predict 
precise values of the underlying physical quantities at a later time. 
This is achieved just by solving the equations of motion. However, 
in the quantum theory, one predicts only the probabilities of possible 
values of the physical quantities, even though one knows the state at 
a given time as precisely as possible. In the wave theory, 
this has been one of the main physical motivation underlying 
behind the hypothesis of Heisenberg uncertainty principle. 
Precisely, we can never achieve a state in which both uncertainties, \textit{viz.} 
$( \Delta x, \Delta p )$ or $( \Delta t, \Delta E )$ vanish identically.
The limit of smallness in these errors obeys the following restriction
$ \Delta x \Delta p \geq h $ or $ \Delta t \Delta E \geq h $.
From the perspective of operator theory, the above notion follows 
from the Heisenberg-Weyl algebra $ \Delta x \Delta p \geq
\vert < [\widehat{x}, \widehat{p}] > \vert $ with $[\widehat{x},
\widehat{p}]:= i \hbar $. On the other hand, for the general case of a family 
of $ L^2 $ functions, one can arrive at the Heisenberg uncertainty principle,
which has been well celebrated since the history of the quantum mechanics,
simply by using the standard deviation of the norm $ \vert f \vert^2 $.
Let $ \Delta (f) $ be  the measure of spread of $ f $ and $ \Delta (\widehat{f}) $
be the corresponding measure of $ \widehat{f} $, then the definition
$ \langle f, g \rangle= \frac{1}{\sqrt{2 \pi}}\int_R \overline{f}(x) g(x) dx $ 
with inner product $ \Vert f \Vert_2^2=  \langle f, f \rangle $ leads to
the Heisenberg uncertainty principle. The argument behind the above 
synthesis follows from the definition of Fourier transformation 
$ \widehat{f}(k)= (\Im f)(k)= \langle e^{isk}, f(s) \rangle $. 
The fact that the centroid of $ \vert f \vert^2 $ and $
\vert \widehat{f} \vert^2 $ may be chosen at the origin yields
$ \Delta (f)= \frac{\Vert xf(x) \Vert_2}{\Vert f(x) \Vert_2} $. Herewith, one
achieves the Heisenberg uncertainty principle as the inequality
$ \Delta (f) \Delta (\widehat{f}) \geq \frac{h}{4 \pi} $.


The above uncertainty relations are one of the absolute feature
of any quantum theory and they resolve the apparent conflict between
the particle and wave natures. This unifies both of them as mutually 
complementary aspects of the physical degrees of freedom.
In the microscopic domain, the nature of the quantum states of 
arbitrary physical processes depends on the background uncertainties.
In this concern, the uncertainty principle states that there always 
exist quantum fluctuations on which we can never have the complete 
control by any means. The above statement remains true even in the
absolute vacuum and at the absolute zero temperature. For example, 
the fluctuations of the energy and momentum can become arbitrarily
large, as we go to the shorter and shorter spacetime intervals.
Of course, the experimental apparatus so far designed till the present date
always have a limitation in their precision of the measurements. This shows 
that we cannot probe the spacetime structure at arbitrarily short intervals. 
Because of the relationship among the precision measurements, 
we may ignore the quantum fluctuations associated with arbitrarily short
intervals, which exceed the order of the precision of a measuring apparatus. 
When we take a quantum mechanical existence of the gravity, the situation is
completely ruined. This is because gravity directly couples to the energy momentum 
tensor, which increases without a definite limit, at arbitrarily short distance scales, 
and thus it directly affects the strength of the gravitational interactions. 
Due to the above reason, when we apply the renormalization technique to 
the general relativity, we are forced to introduce an infinite number of
undetermined constants, which must be fitted to the experimental data.
Therefore, the renormalization method looses its power for the gravity, 
\textit{viz.} the theory of quantum gravity requires an infinite
number of counter terms in order to make its predictions.

On the other hands, in order to determine the physical quantities,
which describe on-mass-shell scattering amplitudes, \textit{e.g.}, $S$-matrix, 
one needs to develop a theory whose ingredients can be deduced from a
quantum field theory. For example, the unitarity and maximal analyticity
conditions of the $S$-matrix encode the requirements of causality and 
non-negativity of the probabilities. In this perspective, 
there are two possible questions, namely, (i) is it only the failure of 
perturbative renormalization theory and not the failure of the
general relativity theory itself? (ii) should the general relativity be
modified at the short distance regimes, irrespective of the validity 
of the perturbation theory, such that the quantum fluctuations of the
energy momentum tensor become large? The significance of the above conflict
between the general relativity and the quantum theory is so profound that
we cannot prevent ourselves from these difficulties in order to concentrate
towards its final resolution. The resolution as by now understood is that
the string theory can be regarded as a sort of the final outcome of many 
essential ideas springing from various attempts towards the fundamental 
theory of the all known interactions. The final answer of both the above 
questions has not yet been fundamentally obtained. However, it is at least 
true from the exploration of the string theory \cite{GreenSchwarzWitten},\cite{Polchinski} 
that we are uncovering a multitude of the facets of the theory. Apart from the general 
understanding of string theories themselves, this could be useful in order to strengthen 
out our understanding of the gauge theory and general relativity in a quite unexpected way. 
The basic reason to make such a framework possible is the fact that both the gauge field theories 
and general relativity are inextricably intertwined as the same facets of the string theory.

\section{Stringy Uncertainty Principle}

In this section, we consider the analysis of stringy $\alpha^{\prime} $ corrections
and their role towards the generalized uncertainty inequalities. From the perspective
of complex analysis, we show that there exists a class of non-zero corrections from both
the holomorphic and anti-holomorphic sectors and their mixing. It is well known
that the facets of string theory automatically contain gravity. Such a
consideration provides a remarkable arena, where various physical ideas and 
the corresponding mathematical structures which are regarded as being entirely
unrelated, can be unified. What follows next that there is a great amount of  
possibility that we can ultimately achieve the unified field theory of nature. 
At this juncture, it appears that the string theory has a set of astonishingly rich
structures, and it possesses many features which are desirable from the perspective 
of the ultimate theory of nature. This follows from the existence of the gravity mode 
in string theory. In this concern, the establishment of Scherk and Schwarz \cite{ScherkSchwarz} 
suggests that the string theory should be regarded as the fundamental theory. Subsequently, 
the string theory has been taken more seriously as the fundamental theory. 
In order to construct a consistent theory of the quantum gravity, the framework of 
string theory receives importance because of various failure attempts of unifying gravity 
with other standard local field theories. Furthermore, the extreme self consistency 
of string theory is not a defect, but it can be interpreted as the 
most important signature towards the ultimate unification.

The string theory \cite{GreenSchwarzWitten}, \cite{Polchinski},
as the most promising candidate for the theory of everything, 
includes gravity and standard gauge like forces. In the low energy limit,
when the length of strings can be ignored, the string theory 
approximately describes various appropriate gauge theories of the
familiar types, \textit{e.g.}, Maxwell's electromagnetism. 
The gravitational interactions contained in the string theory
are described by the low energy limit, as standard supergravity
theories. These configurations with extended symmetry properties
have been eventually constructed as possible generalizations of 
the general relativity. In such attempts, apart from the fundamental
string length, the mathematical structure of the theory indicates that
all the parameters of the theory, including the spacetime geometry itself, 
can in principle be determined from the dynamics of the theory itself. 
The appearance of the critical spacetime dimension(s) can be regarded
as a special case of the general feature of the theory. Unfortunately, 
one does not actually think that the meaning and the content of string theory 
are yet fully grasped, at the present stage of the developments. Although, 
in contrast to the limitations of the perturbative quantum gravity,
the string theory has offered resolution towards the problem of divergences 
associated with numerous earlier attempts of quantizing gravity.
However, the string theory is yet needed to be understood from its 
nonperturbative perspective and its explicit connection with the
quantum theory of everything. This shows how deep string theory
could in general be and how difficult it is to find the really
appropriate mathematical language to formulate the principles
behind the string theory. In experts opinion, we may need a new
mathematical framework in order to satisfactorily describe the
whole content of the string theory, \textit{viz.} the principles 
behind the nonperturbative string theory.


It is known that the fundamental difficulty of divergences of
quantum gravity theories are related to the quantum uncertainties which are
resolved in the string theory. In order to anticipate the basic nature 
of string dynamics, we need to understand simply what a string is.
In the fundamental theory of strings, a sting is simply a one dimensional
extended geometric object, whose energy density along the string renders
to be the universal constant $ \frac{1}{2 \pi \alpha^{\prime} } $.
Here, the new fundamental constant $ \alpha^{\prime} $ characterizes
the underlying string theory \cite{GreenSchwarzWitten}. Thus, the
energy per unit length remains constant as the considered string stretches 
or shrinks. Consequently, the total mass of the string can be determined
by its total length, which means that the length of the string, in the
lowest energy state, is at least classically zero, and thus such 
states are massless. When the above string is quantum mechanically treated, 
the quantum fluctuations have to be taken into account. Therefore,
in the classical sense, one cannot state that the length of the 
lowest energy states of the string is strictly zero. In string theory, 
the massless spin-2 close string states, which behave as the graviton,
are responsible for the universal gravitational force. In the low energy limit,
this exactly coincides with graviton that one expects from the 
quantization of the general relativity. Furthermore, the spin-1 open 
string states coincide with the well-known gauge particles, \textit{e.g.}, 
photon in the quantum electrodynamics and gluons in the quantum
cromodynamics \cite{GreenSchwarzWitten}, \cite{Polchinski}. 
In this sense, the string theory in general contains both the gravity 
and gauge interactions. In particular, even if we start only with
open strings, the consistency of the theory requires that
the closed strings must always coexist with the open strings.
Both the open and closed strings do interact that is 
dictated atleast in the low energy supergravity limit. 
In fact, the string length introduces a regulator into the theory.
In contrast to the Feynmann diagrams of standard gauge theories,
the string interactions appear in such a manner that the divergences
in the corresponding string diagrams are absent in peculiar ways
\cite{GreenSchwarzWitten}, \cite{Polchinski}. The above string interactions
are splitting and joining at the end points of open strings and the
rejoining of two closed strings. At arbitrary points, the interactions
involving the open and closed strings amount to the statement that a sufficiently 
small portion of the underlying worldsheet theory is always dynamically equivalent 
to a segment of the one sheeted plane. In this sense, it is worth mentioning that
the uniformity of the worldsheet is mathematically formulated by a characteristic 
conformal invariance requirement which is intimately connected to the universal
nature of the energy density of constituent strings.

One of the important proposal pertaining to the maximal analyticity 
in angular momentum is subsequently led down due to Chew and Frautschi
\cite{ChewFrautschi1},\cite{ChewFrautschi2}. According to this proposal
\cite{ChewFrautschi1},\cite{ChewFrautschi2}, one can uniquely extend the partial 
wave amplitudes $a_l(s) $ to the analytic function $ a(l,s) $ of $ l $ for a given
set of isolated poles, so called the Regge poles. For a given scattering reaction,
the Mandelstam invariant $ s $ is the square of the invariant energy.
The position of a Regge pole is determined by the Regge trajectory 
$ l= \alpha(s) $. Thus, the physical hadron states are determined 
by the values of $ s $ for which $ l $ takes definite physical values. 
It is worth mentioning that the necessity of branch points in $ l $ plane 
turns out to be associated with the Regge cuts \cite{BardakciMandelstam}. 
Phenomenologically, there have been many new hadrons discovered
in experiments for which the mass squared versus angular momentum plot 
with a given set of fixed values of other quantum numbers shows that 
the Regge trajectories are approximately linear with the common slope 
$ \alpha(s ) = \alpha(0) + \alpha^{\prime} s , \alpha
\sim 1.0 (GeV)^{−2} $. On the basis of the crossing symmetry properties
of the analytically continued scattering amplitudes of such hadronic processes, 
one argues that the exchange of the Regge poles in $t$-channel are well controlled 
at the high energies. For a given fixed momentum transfer, this gives the following asymptotic 
behavior of the physical amplitude $ A(s,t)\sim \beta(t) (\frac{s}{s_0})^{\alpha(t)} $,
where $ s \rightarrow \infty, t< 0 $.

In order to give an improved phenomenological description of decay reactions,
\textit{e.g.}, $ \pi + \pi \rightarrow \pi + \omega $,
$ \omega \rightarrow \pi^{+} +\pi^{0} + \pi{−} $,
an exact analytic formula exhibiting the duality properties 
with respect to the linear Regge trajectories has been designed
by Veneziano \cite{GVeneziano}. The corresponding structure 
arises as the sum of the three Euler beta functions 
$ T= A(s,t)+ A(t,u)+ A(u,s) $ satisfying 
$ A(s,t)= \frac{ \Gamma(- \alpha(s)) \Gamma(- \alpha(t))}{ \Gamma(- \alpha(s)- \alpha(t))} $. 
In the string theory, there are several remarkable discoveries of $N$-particle
generalizations of the above Veneziano formula and the associated Virasoro formula \cite{Shapiro},
\cite{FubiniGordonVeneziano},\cite{FairlieNielsen},\cite{GrossNeveuScherkSchwarz}. 
Such a generalization has a consistent factorization as the spectrum of single particle states.
Subsequently, the string spectrum describes an infinite number of harmonic oscillators 
$ \lbrace a^{\mu}_m \rbrace, \mu:= 1,2,\ldots ,d-1; m=1,2,\ldots $. In the Veneziano case, 
there is a single set of such oscillators and in the Virasoro case, there are two sets of oscillators
\cite{Shapiro},\cite{FubiniGordonVeneziano},\cite{FairlieNielsen},\cite{GrossNeveuScherkSchwarz}.
The above observations lead to the fact that the scattering modes describe 
the relativistic nature of string theory, \textit{viz.} the open strings in Veneziano case
and the closed strings in Virasoro case. Furthermore, Ref. \cite{C.Lovelace} shows that 
the branch points become poles for the parameters $ \alpha(0) = 1 $ and $ d = 26 $. 
These poles are interpreted as closed-string modes in the one-loop open-string amplitude.
Such an interpretation is referred as open-closed string duality principle.

On the other hand, the theory of renormalization group is based on
the fact that it is possible to organize physical phenomena
according to the energy (or distance) scale, \textit{i.e.}, the short
distance physics is not directly affected by the qualitative
features of the long distance physics and vice-versa. This sort of
separation between the UV and IR properties holds for point like 
local quantum field theories. However, there exist interrelations 
between the UV and IR physics for the generalizations, \textit{viz.} 
noncommutative field theory and quantum gravity candidates. In particular,
the string theory perspective can explicitly demonstrate the problem
of UV/IR mixing \cite{SeibergWitten},\cite{ConnesDouglasSchwarz},
\cite{BoerGrassiNieuwenhuizen},\cite{DijkgraafVafa},\cite{DanielaBigatti}.
From the viewpoint of probing the short distance spacetime structures, 
the most decisive direction towards the scales of the distance and
momentum (in the light of the string dynamics) are along the strings 
themselves, where the physical pictures are united as per the properties 
of the string worldsheet. In such a simple model, the corresponding mathematical
properties follow directly from the notion of the complex analysis.
Since a point like standard model does not contain gravity explicitly,
thus the generalized uncertainty principle arises as the consequence of
discretization of the spacetime which may or may not be the property 
of full quantum gravity. We leave this issue open for a future investigation. 
In this concern, it is worth mentioning that a complete explanation of the generalized
uncertainty principle turns out to be useful in understanding realistic physical 
situations. Subsequently, we offer the generalized uncertainty principle in its most
general form and illustrate its role towards the length scales of nature.
In particular, the notion of complex analysis shows that the main theorem
of the present investigation offers a resolution criterion towards the UV/IR mixing problem. 
Further, for a given energy dependent function, our proposition sheds light on both the 
short distance geometries and the possible quantum gravity candidate theories.


In order to discuss the above issues,
let us now turn our attention on the analysis of generalized uncertainty
relations pertaining to the most general quantum mechanical physical system.
For any complex function $ f(x) $, let us first recall the concept of 
$ \epsilon , \delta $ limit and standard theory of Fourier transformation \cite{Rudin}. 
For given $ \epsilon > 0 , \exists \ \delta > 0 $, and there exists $ f(x) $ 
such that $ \vert f(x) - f(x_0)\vert < \epsilon $, whenever $ \vert
x - x_0 \vert < \delta $. In the sense of $ L^2 $ convergence \cite{Rudin}, 
consider a Fourier pair $ (f,\widehat{f}) $ with the Fourier conjugates $(x,k)$ 
satisfying $ f(x)= \frac{1}{2 \pi} \int dk \widehat{f}(k) e^{i k . x} $. 
As shown in section $2$, this implies that we have $ \Delta x = \frac{a}{\Delta k} $, 
where $a$ is a constant, which depends on the size and shape of the wave packet. 
For a given energy scale $\varepsilon $, let us now consider a set of step sizes 
$ \lbrace \varepsilon_i \rbrace_{i=1}^N $ and view this set of energies as a finite 
sequence of real numbers. Consider arbitrary real lattice with a sequence of 
variable step sizes $ \lbrace \varepsilon_i - \varepsilon_j \rbrace $. 
Let $ \epsilon= max_{i,j \in \Lambda} \lbrace \vert \varepsilon_i - \varepsilon_j \vert \rbrace $ 
be the maximum step size and $\Lambda= \lbrace 1,2,\ldots, N \rbrace $ 
be the corresponding index set. This amounts to state that the equation 
$ \Delta x = \frac{a}{\Delta k} $ holds, if $ \epsilon \rightarrow 0 $. 
From the perspective of quantum relativity, the present investigation demonstrates 
that there exist a sequence of functions containing the minimum and maximum length scales.
In this sense, as discussed in section $2$, it is worth mentioning that the concept of scaling 
modifies the standard uncertainty principle of quantum mechanics. From the perspective of 
fundamental string theory \cite{GreenSchwarzWitten},\cite{Polchinski}, the subsequent 
analysis shows that the uncertainty inequalities offer a set of important clues towards 
the physical understanding of the unification problem and short distance geometries.

\subsection*{Proposition (Existence)}

There exist a function $ \tilde{\delta}(\Delta x, \Delta k, \epsilon) $ 
satisfying $ \tilde{\delta}(\Delta x, \Delta k, \epsilon) \rightarrow 0
$ such that

\begin{equation} \label{prop}
\Delta x = \frac{a}{\Delta k} + \tilde{\delta}(\Delta x, \Delta k, \epsilon),
\end{equation}

whenever $ \epsilon \rightarrow 0 $.

\subsection*{Lemma}

For a given real valued function 
$ f(x,y), \ \forall \ x,y \ \in R $, there exists a complex valued function 
$ F(z):= u(x,y) + i v(x,y); z= x+ iy \in C $, where  $ u,v $ are two real functions.
In other words, the map $ f(x,y) \longrightarrow F(z) $ is an isomorphism.

\subsection*{Proof Of The Lamma}

Let $ (x,y) \in R^2 $ be a given ordered pair of $ x,y \in R $,
then we can consider the Cartesian representation of $x$ and $y$
such that $ \forall \ (x,y) $, the pair $ (x,y) $ corresponds to
a real function $ (x,y) \mapsto f(x,y) $. This mapping can be 
represented by the basis vectors $(\widehat{i}, \widehat{j})$
and its general vector as the expansion 
$ \overrightarrow{r}= x \widehat{i} + y \widehat{j} $. 
Now the above consideration in Argand plane implies that 
$ z = x + i y ; x = Re(z), y= Im(z) $. As desired, this amounts to state that
the identification $ z= Re(z)+ i Im(z)= ( Re(z), Im(z) ) $ corresponds
to an uniquely determined $ F(z) $. Hence, $ \forall \ f(x,y), \ \exists \ F(z) $ 
such that (i) $(0,0) \mapsto 0+i0 $, (ii) we have $ (x,y) \in R^2, \ \forall \ z =(x+iy) $
and (iii) $ z =(x+iy) \in C $ is unique, $ \forall \ (x,y) \in R^2 $. 
As per the present consideration, $ F(z) $ is a complex function, 
thus it can further be expressed as $ F(z):= u(x,y) + i v(x,y) $,
where $u(x,y)$ and $v(x,y)$ are two real functions. Hence, the identification
$ f(x,y) \longmapsto F(z) $ is an isomorphism. 

\subsection*{Theorem} \label{alpha} 

There $ \exists \ \alpha^{\prime} $ corrections,
iff the function $ \tilde{\delta}(\Delta x, \Delta k, \epsilon) $ 
factorizes either into an analytic function with respect to $ (\Delta k, \epsilon) $
or an anti-analytic function with respect to $ (\Delta x, \epsilon) $.

\subsection*{Proof Of The Theorem} 

Assuming that $ \exists \ \tilde{\delta}(\Delta x, \Delta k, \epsilon) $ such that

\begin{equation} \label{theorem}
lim_{\epsilon \rightarrow 0} \tilde{\delta}(\Delta x, \Delta k, \epsilon) \longrightarrow 0.
\end{equation}

From the Eqn.(\ref{prop}), the vanishing of $ \tilde{\delta}(\Delta x, \Delta k, \epsilon) $
implies that the one of the uncertainties $\{ \Delta x , \Delta k \}$ can be uniquely determined from the other.
In general, the variables $\{ \Delta x , \Delta k \}$ depend on the fundamental energy parameter $ \epsilon$. 
Subsequently, we see that $ \Delta x $ can be represented in terms of $ \Delta k $ and $ \epsilon $. 
Hence, it is enough to consider a real reduced function $ \tilde{\tilde{\delta}}(\Delta k, \epsilon) = C
\tilde{\delta}(\Delta x, \Delta k, \epsilon) $, where $C$ is a constant. 
Without ambiguity, it follows from the Eqn.(\ref{theorem})  that we have 
$ \tilde{\tilde{\delta }}(\Delta k, \epsilon) \longrightarrow 0 $, whenever
$ \epsilon \rightarrow 0 $.

From the complex function theory, we know for any complex function
$ \tilde{\tilde{\tilde{\delta }}} $ on an open set $ U $ with $
\tilde{\tilde{\tilde{\delta }}}(\Delta k+ i \epsilon)= u(\Delta k,
\epsilon)+ iv((\Delta k, \epsilon)) $, where the real functions $
u(\Delta k, \epsilon) $ and $ v(\Delta k, \epsilon) $ are real and
imaginary parts of $ \tilde{\tilde{\tilde{\delta }}} $ that
the differentiability condition of $ \tilde{\tilde{\tilde{\delta
}}} $ are expressed in terms of $ u(\Delta k, \epsilon) $ and $
v(\Delta k, \epsilon) $. To do so, let us concentrate on the holomorphic
sector and derive the corresponding equivalent conditions towards the
differentiability of $ \tilde{\tilde{\tilde{\delta }}} $ as a function
of the $ u(\Delta k, \epsilon) $ and $ v(\Delta k, \epsilon) $. For a fix 
$ z $, let $ \tilde{\tilde{\tilde{\delta^{\prime}}}}(z)= a_1+ i a_2
$ and $ w= h_1 + ih_2; h_1,h_2 \in R $. Now, suppose that $
\tilde{\tilde{\tilde{\delta }}}(z+w)- \tilde{\tilde{\tilde{\delta
}}}(z)= \tilde{\tilde{\tilde{\delta^{\prime} }}}(z)w + \sigma(w)w
$ with $ Lim_{w \rightarrow 0} \sigma(w)= 0 $. Then, it follows that we have 
$\tilde{\tilde{\tilde{\delta^{\prime} }}}(z)w= (a_1+ ia_2)(h_1+
ih_2)= a_1 h_1- a_2 h_2 +i(a_2 h_1+ a_1 h_2)$.
On the other hand, let $ \tilde{\tilde{\delta}}:U \rightarrow R^2 $ be
a vector field such that $ \tilde{\tilde{\delta}}(x,y)= (u(\Delta
k, \epsilon), v(\Delta k, \epsilon)) $. In this case, we emphasize 
the fact that $\tilde{\tilde{\delta}} $ is a real vector field associated 
with the function $\tilde{\tilde{\tilde{\delta}}} $. With the condition 
$ \sigma_j(h_1,h_2)_{(h_1,h_2)\rightarrow 0}\rightarrow 0 $,
it follows that the function $ \tilde{\tilde{\delta}}$ satisfies
the expansion $ \tilde{\tilde{\delta}}(\Delta k+h_1,\epsilon+h_2)-
\tilde{\tilde{\delta}}(\Delta k,\epsilon)= a_1 h_1- a_2 h_2 +i(a_2
h_1+ a_1 h_2) + \sigma_1(h_1,h_2)h_1+ \sigma_2(h_1,h_2)h_2 $, where $ j= 1,2 $.

Assuming that the function $ \tilde{\tilde{\tilde{\delta }}} $ is holomorphic,
thence one concludes that the corresponding $ \tilde{\tilde{\delta}} $ is
differentiable in the sense of real variables. In particular, the derivatives
can be represented by the associated Jacobian matrix 
$ J_{\tilde{\tilde{\delta}}}(\Delta k,\epsilon)= \left
(\begin{array}{rr}
    a & -b \\
     b & a \\
\end{array} \right)
=\left (\begin{array}{rr}
    u_{\Delta k} & -u_\epsilon \\
     v_{\Delta k} & v_\epsilon \\
\end{array} \right)
\Rightarrow \tilde{\tilde{\tilde{\delta^{\prime} }}}(z)=
\frac{\partial u}{\partial \Delta k}- i \frac{\partial v}{\partial
\epsilon} $ and $ u_{\Delta k}= v_\epsilon= a; u_\epsilon=
-v_{\Delta k}= -b $. The equivalence is described by the celebrated Cauchy-Riemann
equations. Conversely, let $ u(\Delta k,\epsilon) $ and $ v(\Delta
k,\epsilon) $ be two continuously differentiable real valued functions
satisfying the above Cauchy Riemann equations. Then, in order to proceed further,
let us define a complex valued function $\tilde{\tilde{\tilde{\delta }}}(z)= \tilde{\tilde{\tilde{\delta
}}} (\Delta k+i\epsilon)=  u(\Delta k,\epsilon)+ iv(\Delta k,\epsilon) $.
Herewith, by reversing the above steps, it immediately follows that 
$ \tilde{\tilde{\tilde{\delta }}} $ is differentiable in the sense of
complex variables, \textit{i.e.}, $ \tilde{\tilde{\tilde{\delta }}} $ is
a holomorphic function. Notice further that the above Jacobian determinant 
$ J_{\tilde{\tilde{\delta}}}(\Delta k,\epsilon)$ can be expressed as
$\Delta_{\tilde{\tilde{\delta}}}= a^2+ b^2= u_{\Delta k}^2 +
v_{\Delta k}^2= u_{\Delta k}^2+ u_\epsilon^2 $. Thus, we have the condition that
$ \Delta_{\tilde{\tilde{\delta}}} \geq 0 $. Interestingly, it turns out that
$ \Delta_{\tilde{\tilde{\delta}}} $ is non zero, iff 
$ \tilde{\tilde{\tilde{\delta^{\prime} }}}(z) \neq 0 $. 

Herewith, we observe that the relation between the function
$ \tilde{\tilde{\tilde{\delta^{\prime} }}}(z) $ and the respective
Jacobian determinant of $ \tilde{\tilde{\delta}}$ satisfies
$ \Delta_{\tilde{\tilde{\delta}}}(\Delta k,\epsilon)= \vert
\tilde{\tilde{\tilde{\delta^{\prime} }}}(z) \vert^2 $. 
On the other hand, the above equation with the definition 
$ \tilde{\tilde{\tilde{\delta}}} (\Delta k, \epsilon)= u(\Delta k,
\epsilon) + i v(\Delta k, \epsilon) $ implies that $
u(\Delta k, \epsilon)\rightarrow 0 $ and $ v(\Delta k,
\epsilon)\rightarrow 0 $, whenever $ \epsilon \rightarrow 0 $.
Conversely, this amounts to state that, 
if Cauchy- Riemann conditions are satisfied, then 
$\tilde{\tilde{\delta }}(\Delta k, \epsilon) $ is an analytic function. 
Whence, we have $ u_{\Delta k}= v_{\epsilon} $ and $ u_{\epsilon}= - v_{\Delta k}
$. Since $ u(\Delta k, \epsilon), v(\Delta k, \epsilon)\rightarrow
0 $ as $ \epsilon \rightarrow 0 $, thus $ v_{\epsilon} $ and $
u_{\epsilon} $ do exit. As mentioned above, the Cauchy-Riemann equations
imply that $ u_{\Delta k} $ and $ v_{\Delta k} $ exit, as well. 
Consequently, there exists an analytic function $ \tilde{\tilde{\delta }}(\Delta k,
\epsilon) $ and thus the $ \alpha^{\prime} $ corrections. Furthermore,
for the case of an anti-analytic function, it is worth mentioning that 
a similar examination leads to the anti-Cauchy-Riemann conditions.

\subsection*{Proof Of The Proposition}

Let the function $ \tilde{\tilde{\delta}}(\Delta k, \epsilon) $ be considered
as an analytic map. Then, the above theorem with the Eqn.(\ref{theorem}) implies 
that we have $ \tilde{\tilde{\delta}}(\Delta k, \epsilon) =  
\tilde{\tilde{\delta}}(0,0) + \Delta k \frac{\partial
\tilde{\tilde{\delta}}}{\partial \Delta k} \vert_0 + \Delta
\varepsilon \frac{\partial \tilde{ \tilde{\delta}}}{ \partial
\Delta \varepsilon} \vert_0 +\ldots $ in the sense of Taylor
series of a real valued function.
For instance, in the case of a single variable real valued function,
it amounts to the statement that $ f(\xi)= f(\xi_0)+ (\xi -
\xi_0) \frac{df(\xi)}{d\xi}\vert_{\xi= \xi_0} + \ldots $. From
the Eqn.(\ref{theorem}), we have $ \Delta x = \frac{a}{\Delta k} +
\frac{\tilde{ \tilde{\delta}}(0,0)}{C} + \frac{\alpha}{C}\Delta k
+ \frac{\gamma}{C}\Delta \varepsilon +\ldots $, where $ \alpha:=
\frac{\partial \tilde{\tilde{\delta}}}{ \partial \Delta k} \vert_0
$ and $ \gamma:= \frac{\partial \tilde{\tilde{\delta}}}{ \partial
\Delta \epsilon}\vert_0 $. This yields that $ \Delta x = \frac{a}{\Delta k} +
\frac{\alpha}{C} \Delta k + \frac{\gamma}{C} \Delta \varepsilon +
\frac{\tilde{\tilde{\delta}}(0)}{C}+ \ldots $. In the above expansion,
$ \tilde{\tilde{\delta}}(0) $ is kept arbitrarily small constant, thus
it may be scaled down to zero as the limit, \textit{viz.} we have $ \tilde{\tilde{\delta}}(0) =0 $.
Similarly, for a suitable set $ \lbrace \varepsilon_i \rbrace $, we can 
chose a sequence $\varepsilon $ such that $ \Delta \varepsilon = 0 $.
Physically, such a sequence corresponds to the case of an equispaced real lattice. 

On the other hand, the first order approximation of $ \tilde{\tilde{\delta}}$
implies that we have $ \Delta x= \frac{a}{\Delta k} + \frac{\alpha}{C} \Delta k $. 
From the dilation $ k \rightarrow a k $, we can define a transformed $ x $ such that the
Fourier transform of the aforementioned function $ f(x) $ remains the same. 
At this order of analysis, the normalization of $f(x)$ results into the scaling 
$ \Delta x = \frac{1}{\Delta k} + \tilde{\alpha} \Delta k $. Define a new variable 
$ p = \hbar k $, then we have $ \Delta x = \frac{\hbar}{\Delta p}
+ \frac{\tilde{\alpha}}{\hbar} \Delta p $, 
where $ \tilde{\alpha}:= \frac{\alpha a}{C} $. 
In other words, at the leading order contributions,
we have $ \tilde{\delta}(\Delta x, \Delta k, \epsilon)=
C \tilde{\alpha} \Delta k $. Furthermore, our analysis can be easily
generalized to the case of a non-zero $ \tilde{\tilde{\delta}}(0) $,
\textit{viz.} $ \tilde{\tilde{\delta}}(0) \neq 0 $. In this case,
we find that the equality relation is modified to the following inequality
$ \Delta x \geq \frac{\hbar}{\Delta p} + \frac{\tilde{\alpha}}{\hbar} \Delta p $.
It is obvious that, even if $ \Delta \varepsilon \neq 0 $, 
yet such a generalized uncertainty principle holds. 
This completes the perturbative proof of the proposition.
In the next section, we illustrate that the nonperturbative
properties follow from the duality symmetries of the theory.

\subsection*{Remark Cum Corollary}

For arbitrary quantum physical situation, we notice via the notion of $T$-duality
that the nonperturbative spacetime corrections are contained in the function 
$ \tilde{\delta}(\Delta x, \Delta k, \epsilon) $.  The perturbative 
corrections come into the picture, whenever there exist the two 
variable factorizations, \textit{i.e.}, (i) $ \tilde{\delta}(\Delta x, \Delta k,
\epsilon)= C_1(\Delta x) \tilde{\tilde{\delta}}( \Delta k,
\epsilon) $ and (ii) $ \tilde{\delta} (\Delta x, \Delta k, \epsilon)=
C_2(\Delta k)\tilde{\tilde{\delta}}_1( \Delta x, \epsilon) $. 
In this sense, our analysis of the generalized uncertainty principle
yields a resolution criteria towards the UV/IR mixing problem.
In general, it is worth mentioning that the functions 
$ \tilde{\delta}(\Delta x,\Delta k, \epsilon) $ contains suitable 
cutoffs operating at all order of the energy (or length) scale. 
Moreover, we observe that the first Taylor coefficient of 
$ \tilde{\tilde{\delta}}( \Delta k, \epsilon) $ determines 
various physical quantities operating in gauge theories,
quantum gravity candidates and unified theories. For example,
an interesting instance arises from the first perturbative
coefficient of $ \tilde{\delta}(\Delta x,\Delta k, \epsilon)$.
In this concern, the Planck length scale and string length scale
can be determined as $ \tilde{\alpha}= \alpha^{\prime} l_p^2= l_s^2 $.

\subsection*{Generalization}

In a more realistic physical situation, the generalized 
uncertainty principle arises as per its following extension.
Consider a general $ d $ dimensional manifold $ (\mathcal M, g) $
and arbitrary sequence $ \lbrace \varepsilon_i \ \vert \ i \in \Lambda \rbrace $
with $ \varepsilon_i:= (\varepsilon_i^1,\varepsilon_i^2, \ldots , \varepsilon_i^d) \in
\mathcal M $. Define $L^2$ norm as $ \Vert \varepsilon_i-
\varepsilon_j \Vert_2^2:= \sum_{\alpha,\beta=1}^d g_{\alpha \beta}
( \varepsilon^{\alpha}_i- \varepsilon^{\alpha}_j)
(\varepsilon^{\beta}_i- \varepsilon^{\beta}_j) $, where $ d=
dim(\mathcal M) $. Then, for a fixed $ j \in \Lambda $, 
the various step sizes form the following sequence
$ \lbrace \epsilon_i:= \Vert \varepsilon_i- \varepsilon_j \Vert \rbrace $,
where $ i \in \Lambda $. For a given index set $ \Lambda $, 
let $ \epsilon:= max_{i,j \in \Lambda} \lbrace \Vert \varepsilon_i- 
\varepsilon_j \Vert \rbrace $ be the maximum step size.
Then $ \forall \ i,j \in \Lambda $, we see that $ \epsilon \geq \Vert
\varepsilon_i- \varepsilon_j \Vert $. This is a local definition of $ \epsilon $, 
which is in perfect agreement with standard $d$-dimensional $ R^d$.
With the notation $ g_{\alpha \beta}= \delta_{\alpha \beta} $, 
we find that the case of $R^d$ arises with the norm $ \Vert 
\varepsilon_i- \varepsilon_j \Vert_2= \sqrt{(\varepsilon_i^1-
\varepsilon_j^1)^2+ (\varepsilon_i^2- \varepsilon_j^2)^2+ \ldots +
(\varepsilon_i^d- \varepsilon_j^d)^2} $. For one dimensional situations,
it reduces to $ \epsilon:= max_{i,j \in \Lambda} \lbrace \vert \varepsilon_i-
\varepsilon_j \vert \rbrace$. It is worth mentioning that the above metric is 
defined as the two norm $ d \varepsilon^2:= \sum_{\alpha,\beta=1}^d g_{\alpha \beta} d
\varepsilon^{\alpha} d \varepsilon^{\beta} \leq \epsilon^2 $.
Notice further that $\epsilon= 0 $ iff $ \varepsilon_i= \varepsilon_j, \forall \ i,j \in \Lambda $. 
In particular, for any general $d$-dimensional lattice $ Z^d $, 
we restrict $ \varepsilon_i $ to $ Z^d $. Therefore, 
for a given $ \Lambda $, we  arrive at the following two cases 
(i) when $ \Lambda $ is a finite set, the underlying manifold 
$ \mathcal M $ can be finitely covered by a suitable collection of open sets $ U_i $
and so $ \mathcal M$ is compact with $ \mathcal M= \bigcup_{i \in \Lambda} U_i $,
(ii) when $ \Lambda $ is infinite set, then the manifold $ \mathcal M $ cannot be 
finitely covered and thus $\mathcal{M}$ renders to be a non compact manifold. 
In both the above cases, we find that there exists a function 
$ \tilde{\delta}(\Delta \overrightarrow{x}, \Delta \overrightarrow{k}, \varepsilon_i) $
such that the generalized uncertainty principle holds, as given in Eqn.(\ref{prop})
for the special case of $ d= 1 $. In the next section, we shall illustrated that 
the function $\tilde{\delta}(\Delta \overrightarrow{x}, \Delta \overrightarrow{k}, \epsilon) $ 
contains in general all possible stringy $ \alpha^{\prime} $ corrections, UV/IR mixing
peroperties, and the minimum and maximum length scales originating from the perspective
of quantum spacetime and thereby the noncommutative nature of short distance geometries.


\section{The function $ \tilde{\delta}(\Delta x,\Delta k, \epsilon) $}

In this section, we examine a set of generic properties of the 
function $ \tilde{\delta}(\Delta x,\Delta k, \epsilon) $ from
the known physical and mathematical results. 
As mentioned in the foregoing section, it is worth recalling
that our results have been largely based on the analysis of
the function $ \tilde{\delta}(\Delta x,\Delta k, \epsilon) $.
In the sequel, we compare the undermined consequences from
the perspective of important recent developments in the subject,
\textit{viz.} quantum gravity effects, black hole physics,
existence of the maximum/ minimum length scales, string theory, 
short distance geometries, Fourier transformation, distribution theory 
and discretization of the physical spacetime. As mentioned in the 
previous section, we shall show that the present analysis is in 
perfect accordance with the above developments. Such issues
give rise to a set of physical and mathematical indications towards
the appropriateness of the short distance geometries and unification theories. 
In the above perspectives, we show from the consideration of the section $3$
that the underlying properties of $ \tilde{\delta}(\Delta x,\Delta k, \epsilon) $ 
lead a set of new mathematical interests and physical relevance to the 
completely generalized uncertainty principle.

\subsection{Quantum Gravity Effects}

The quantum mechanical uncertainty relations, as the limiting properties
of $\tilde{\delta}(\Delta x, \Delta k, \epsilon) $, are useful towards
the understanding of a class of qualitative explanations of short distance
divergences pertaining to the quantum theory of gravity. Notice that
the string theory unifies the force of gravity with the other gauge forces
and resolves the difficulty of divergences. This exhibits a set of promising
new structures, which could not be envisaged if one limits the consideration
within the framework of standard local point like field theories. In this viewpoint,
the function $ \tilde{\delta}(\Delta x, \Delta k, \epsilon) $ leads to the most
general uncertainty relation. For example, the time energy uncertainty relation,
as formulated by reinterpreting the spacetime distance scales, resolves the
divergence problem. It is worth mentioning that $ \tilde{\delta}(\Delta x, \Delta k, \epsilon) $
reflects the existence of the minimum length scale on a given $(n+1)$-dimensional spacetime manifold.
Such a proposal leads to a new uncertainty relation, which is consistent
with a discretization of the spacetime \cite{CarloRovelli}. At the Planck scale,
this requires a replacement of the standard Lorentzian group symmetries by 
the corresponding quantum group symmetries. This follows from the fact that
the small scale spacetime involves other symmetries than the 
familiar Poincar\'e symmetries.


From nonpertubative quantum string dynamics, the term emerging 
from the classical gravity theory, \textit{viz.}  
$ \Delta x^{ \mu} \geq \Delta p^{\mu} $ indicates a class of 
long distance effects \cite{DGrossPFMende},\cite{KKonishiGPautiPProvero},
\cite{DAmatiMCiafaloniandGVeneziano},\cite{AmatiMCiafaloniandGVeneziano2},
\cite{MFabrichesiGVeneziano},\cite{RGuidaKKonishiPProvero}.
Consequently, the scattering of a high energy particle is dominated by an
exchange of the gravitons \cite{GtHooft}. Thus, the energies higher than
the rest mass energy $ m_p c^2 $ produce a black hole and thereby they
lead to coherent emissions of the real gravitons. On the other hand, the regimes,
where $ \Delta x^{\mu} \geq \frac{1}{\Delta p^{\mu}} $, are characterized
by ordinary gauge interactions. As per the notion of standard 
quantum mechanics, there are two physical regimes which are the same
manifestation of the quantum string dynamics. These regimes are characterized
by the energy scale of the theory, \textit{viz.} the one where the chosen
scales are much larger than the Planck energy, and the other where the scales
are much smaller than the Planck energy.
Within the context of the scale relativity, Castro has led 
an interesting proposal that $ \Delta x^{\mu} \geq {\Delta p^{\mu}} $
behavior originates from an existence of the upper scale in nature. 
This arises from the long distance effects concerning the classical
gravity interactions at the cosmological scale.
Whilst, the term $ \Delta x^{\mu} \geq \frac{1}{\Delta p^{\mu}} $  
results from the standard quantum mechanical origin of fractal
structure of the spacetime at microphysical scales
\cite{CarlosCastro},\cite{LNottaleIJMP1},\cite{LNottaleIJMP2},\cite{LNottale}. 
Hereby, our analysis demonstrates that both the above physical situations can be 
described by the single function $ \tilde{\delta}(\Delta x, \Delta k, \epsilon) $. 
In this fashion, the dilations of $ \tilde{\delta}(\Delta x, \Delta k, \epsilon)
$ merge with the string theory, where the upper length scale $ L $ could
have an interesting functional relation with the Planck scale physics.
As per the Nottale’s proposal \cite{LNottaleIJMP1},\cite{LNottaleIJMP2},\cite{LNottale}, 
our consideration of the completely generalized uncertainty principle 
leads a resolution towards the cosmological constant problem.

The function $ \tilde{\delta}(\Delta x, \Delta k, \epsilon) $ resolves the UV/IR 
mixing problem in the cases when our remark given in the previous section on the factorization 
of $ \tilde{\delta}(\Delta x, \Delta k, \epsilon) $ holds. In this case,
we find the existence of perturbative stringy $ \alpha^{\prime} $ corrections.
Moreover, the string theory relationship is realized as the $T$-duality symmetry,
\textit{viz.} $ R \leftrightarrow \frac{\alpha^{\prime}}{R} $. At this length
scale, the string theory does not distinguish a large radius spacetime
from the dual small radius spacetime. In this concern, Ref. \cite{AshokeSen} 
offers an interesting review of nonperturbative symmetries in string theory. 
On the other hand, the fractal structure of spacetime manifold turns out to be
a crucial question in any theory of the quantum gravity.
Subsequently, the notion of scaling properties can be generalized for
certain non abelian discrete groups $ \Gamma $. In the geometric
model of gravity, this is achieved by considering a set of equivalence 
classes of the unitary irreducible representations $ \widehat{\Gamma} $
of a finitely generated discrete group $ \Gamma $. 
From the viewpoints of measure theory, the algebraic structures
of the corresponding dual $ \widehat{\Gamma} $ are governed by the 
dimension $d$ of the manifold and the number of the underlying generators
$ r $ of the $ \Gamma $. In contrast to the standard Schr\"odinger operators,
such structures can be independently analyzed from the one and the other.
In an abstract geometric situation, the properties of the $ \widehat{\Gamma} $
show that there exists a set of generating spectral gaps in the theory. 
In particular, let $ \mathcal N $ be a non-compact Riemannian covering
manifold and $ \Gamma $ be the corresponding discrete isometry group,
then there exists the Laplacian $ \Delta_{ \mathcal N } $ such that the
quotient $ \frac{ \mathcal N}{\Gamma} $ is a compact manifold. 
For a suitable class of manifolds $ \mathcal N $ possessing a 
nonabelian covering transformation groups $ \Gamma $, the principle
of the minimum and maximum scales implies that there exists a set of 
finitely many gaps in the spectra of the operator $ \Delta_{\mathcal N} $
\cite{FernandoPost1},\cite{FernandoPost2}. In this way, 
the quantum mechanical behavior is obtained by the functional average 
over all possible equivalence classes of the metrices. Consequently,
such a consideration leads to the notion of an effective dimension of the spacetime. 
As introduced in Refs. \cite{LNottaleIJMP1},\cite{LNottaleIJMP2},\cite{LNottale},
the corresponding quantum mechanical measurement arises from 
the fractal nature of the particle trajectories.

\subsection{Black Holes And Existence Of The Minimal Length Scale}

As per the notions introduced in the foregoing section, 
we now wish to address the question \textit{``why the minimum scale 
must appear in the fundamental equation of nature?''}. As proposed by 
Susskind \cite{LSusskind}, the size of strings increases instead of 
decreasing at the energy scales higher than the Planck energy.
Thus, the energy imparted to strings is used to break a string into 
its pieces, \textit{viz.} the building blocks of the vacuum string theory
configuration. Such a consideration leads to the statement of the 
Bekenstein-Hawking bound for the horizon area of a black hole 
\cite{BH1},\cite{BH2},\cite{BH3},\cite{BardeenCarterHawking}.
In this limit, the corresponding entropy of the black hole is given as
$ S= \frac{A}{4G}$. Physically, this amounts to state that one cannot 
have more than one bit of information per unit area in the Planck units. 
This ensures that the uncertainty principle plays an important role 
towards the understanding of the fundamental principles, \textit{viz.}
the formulation of the string theory \cite{GreenSchwarzWitten}, \cite{Polchinski} 
yields the minimum length in nature. The argument behind the above principle follows 
from the fact that the nonpertubative quantum string dynamics pertains to the 
non-existence of a black hole with mass smaller than $2M$ and horizon radius smaller 
than the string length scale $l_s$. Subsequently, the Hawking radiation must stop at the point,
when the Hawking temperature of the black hole reaches to the associated 
Hagedorn temperature $ T = \frac{1}{8 \pi M }= T_H $.  

Ref. \cite{AntalJevickiTamiakiYoneya} suggests that the conformal invariance
of the underlying microscopic theory breaks down at the Hagedorn temperature.
Thus, this is the point of interest after which there is no consistent 
propagation of fundamental strings. In this sense, we find that the 
function $ \tilde{\alpha} $ gives rise to the minimum length parameter 
$ l_p $ of nature as $ \tilde{\alpha}=  \alpha^{\prime} l_p^2 $. 
Notice further that the standard Lorentz transformation properties
do not apply in the world of Planck scale physics. This is because
of the fact that the Riemannian continuum is recaptured only at the 
large distance scales. From the viewpoint of string theory, the notion
of the minimum scale is carried out by the fundamentals of Finsler geometry
\cite{BaoChernShen}. In contrast to the usual Riemannian geometry, the line element 
on a Finsler manifold need not necessarily be a quadratic form \cite{BaoChernShen}. 
As mentioned before, the function $ \tilde{\delta}(\Delta x, \Delta k, \epsilon) $ 
not only reproduces the ordinarily generalized uncertainty relations,
but it yields all possible perturbative and nonperturbative 
stringy corrections operating in nature. In a single stroke, 
this shows a positive sign that the analysis of $ \tilde{\delta} $ 
is on the right track in order to reveal the geometric foundations 
of fundamental theory, \textit{e.g.}, string theory and $M$-theory. 
Furthermore, the properties of Regge trajectories in string theory 
and black hole horizon area quantization follow from the limiting
behavior of $ \tilde{\alpha} $. From the perspective of black hole 
conformal field theories, this opens a new avenue to study the 
microstate counting problem and thus the formation of a black hole.

A priori, Ref. \cite{KouroshNozariandSHamidMehdipour} invokes 
the notion of the black hole thermodynamics with an introduction
of the generalized uncertainty principle. The above analysis shows
that the generalized uncertainty principle modifies the standard black hole 
thermodynamics. In specific, the small black holes emit black body radiation 
\cite{KouroshNozariandSHamidMehdipour} with the following Hawking temperature 
$ T= \frac{Mc^2}{4 \pi}[1 \pm \sqrt{1- \frac{M_p^2}{M^2}}] $.
Here, the minus sign in above expression of temperature $ T $
indicates the large mass limit. In this case, the corresponding
Benkenstein-Hawking entropy \cite{KouroshNozariandSHamidMehdipour} 
is modified as $ S= 2 \pi [\frac{M^2}{M_p^2}(1- \frac{M_p^2}{M^2}+ 
\sqrt{1- \frac{M_p^2}{M^2}})- \ln{\frac{M+ \sqrt{M^2- M_p^2}}{M_p}}]$.
In such a framework of the generalized uncertainty principle, a black 
hole evaporates until its mass reaches the limit of the Plank mass. 
Thus, this viewpoint indicates that the black hole remnants are stable objects.
Such an application of the generalized uncertainty principle to the
black hole thermodynamics suggests a strong possibility of the existence
of black hole remnants. Notice further that the remnant entropy of a black 
hole remnant can be related to fluctuations pertaining to the background 
spacetime metric tensor. For the case of the hydrogen atom and a black hole, 
the generalized uncertainty principle consideration 
\cite{KouroshNozariandSHamidMehdipour} indicates that the hydrogen atom
is unstable and totally collapses, however the black hole evaporates,
until its mass reaches the limit of the Plank mass. The issue
of stability of the remnants can be further considered in the
framework of the symmetry principle. In particular, the supergravity 
consideration provides an important framework towards the understanding
of black hole remnants. According to the above notion, 
the black hole evaporation would terminate when it reduces to a remnant.
This is because of the fact that the graviton spectrum would have
a cutoff at the Plank mass, which at the present should have an
approximate red shifted of $\sim 10^{14} $. Consequently,
we can only measure the position of an object only up to the Plank length $l_p$.
The stringy uncertainty principle leads to the notion that
one cannot set up a measurement to find more accurate
position of the particle than the Planck length itself.
This means that the notion of the locality breaks down at $ l_p $.
As mentioned in the foregoing section, such a behavior of the
particle dynamics is in complete agreement with the functional 
properties of $ \tilde{\delta}(\Delta x, \Delta k, \epsilon) $.
In this sense, the string equations of motion are compatible with
effective physical dynamics than the further generalization of the right
hand side of the generalized uncertainty equation. For example,
in the notations of the previous section, consider a physical 
situation with $ \Delta x= \overline{\lambda}= \frac{\lambda}{2 \pi} $,
$ \Delta p= \hbar \Delta k $, $ \overline{\lambda}= \frac{1}{\Delta k} 
+ \tilde{\alpha} \Delta k $ and $ \omega= \frac{c}{\overline{\lambda}} $,
then the dispersion relation takes the following form 
$ \omega= \omega(k)= \frac{k c}{ (1+ \tilde{\alpha} k^2)} $.


\subsection{String Theory And Short Distance Geometries} 

From the perspective of quantum string theories, 
the string dynamics obeys standard rules of quantum mechanics.
Thus, it is natural to seek a simple universal characterization 
of the stringy properties with respect to the spacetime distance scales.
This can be achieved by invoking the uncertainty relations and thus by 
taking the special characteristics of string theory into the account. 
Notice that the coordinate of the center of mass
and the corresponding momentum of the strings satisfy standard
quantum mechanical coordinate-momentum uncertainty relation, however
the crucial role of the extendedness of strings can be properly 
examined via the energy-time uncertainty relation. This is because
the energy-time uncertainty relation holds in arbitrary dynamical
processes. Specifically, let the uncertainty with respect to the time
is $ \Delta t $, then the energy of a string can be defined in terms
of the length of the string measured along the direction of the string. 
Due to the universal nature of the constant energy density, we may
naturally identify the precision energy scale $ \Delta E $ of an
extension $ \Delta x $ of the longitudinal direction of string
\cite{GreenSchwarzWitten} as per the estimation
$ \Delta E \sim \frac{\Delta x}{\tilde{\alpha }} $.
Therefore, the time-energy uncertainty relation can be
reinterpreted as the following spacetime uncertainty relation
$ \Delta t \Delta x \geq \tilde{\alpha} $, where the new parameter 
$ \tilde{\alpha} $ is defined as the square of the string length 
scale $ l_s $, \textit{viz.} we have $ l_s^2 = \tilde{\alpha} $. 
In this sense, the function $ \tilde{\delta}(\Delta x, \Delta
k, \epsilon) $ provides a precise determination of the stringy 
corrections. Herewith, our completely generalized uncertainty
principle contains all possible perturbative and nonperturbative 
stringy corrections. 

For a non-zero $\tilde{\delta}(\Delta x, \Delta k, \epsilon) $, 
the corresponding spacetime uncertainty relation indicates that the
strings cannot simultaneously probe both the time like and the space like
short distance scales to arbitrary precisions. This follows from the fact
that the string theory is a self contained theory, and thus the spacetime 
structures should be self consistently determined by the dynamics of strings. 
The stringy uncertainty relation qualitatively characterizes the spacetime,
if we consider the fact that the string theory unifies all fundamental 
forces of nature. Indeed, this situation may be expressed by the claim that the
physical spacetime must be `quantized' at the short distances.
In addition to the above, the proportionality between the energy and the
longitudinal length indicates that the large quantum fluctuations
of the energies, which are associated with the short time measurement,
can actually be reinterpreted as the fluctuations of long strings.
Hence, such fluctuations turn out to be a long distance phenomenon. 
From the perspective of the duality principle \cite{AshokeSen}, the structure 
of the quantum fluctuations of string theory is drastically different 
from ordinary local quantum field theory of point particles.
This is precisely the physical mechanism hidden behind the proof that 
the string perturbation theory has no divergences at short distance
spacetime structures \cite{GreenSchwarzWitten},\cite{Polchinski}.


\subsection{Fourier Transformation And Distributions}

Now we focus our attention on the relationship between the theory of
Fourier transformation and a class of generic uncertainty principles.
In a very general perspective, the classical Fourier transformation 
geometrically arises as a duality map between the functions on a vector 
space $ V $ and on its dual vector space $ V^{*} $. For $V= R^d $,
the Fourier transformation can be expressed as the map
$ f(\overrightarrow{x}) \rightarrow \widehat{f}
(\overrightarrow{k})= \frac{1}{(2\pi)^d}\int_V
f(\overrightarrow{x})e^ {i <\overrightarrow{x} \vert
\overrightarrow{k}>} d\overrightarrow{x} $. 
Thus, from the viewpoint of Fourier transform, 
the uncertainty principle asserts that the function $ f $ and its 
Fourier transform $ \widehat{f} $ cannot both be sharply localized. 
For instance, let $ E, \Sigma \subset R^d $ be two measurable sets,
then the definition that (i) $ (E, \Sigma) $ is a `weakly annihilating pair',
if the function $ f \in L^2(R^d) $ vanishes as long as the function
$ f $ is supported in $ E $ and the $ \widehat{f} $ is supported in
$ \Sigma $ and (ii) $ (E, \Sigma) $ is a `strongly annihilating pair',
if $ \exists \ C $ such that, $ \forall \ f \in L^2(R^d) $, the corresponding spectrum 
satisfies $ \Sigma:  \Vert f \Vert_2^2 \leq C \int_{E^C} \vert f \vert^2 dx $. 
Thus, the methods of complex analysis can be applied in order to obtain 
the sharp results on $ E $. In terms of the density, the complement determines 
a precise estimate of the constant $ C $. For example, Ref. \cite{FNazarov}
states that $ E $ and $ \Sigma $ posses finite measure in the case of one dimension.
Ref. \cite{OKovrijkine} indicates that $ \Sigma $ can be viewed as a finite 
union of parallelopipeds. For a given set of annihilating pairs, let us 
first recall the definition of $ \varepsilon $-thin set in $ R^d $. 
Let $ 0 < \varepsilon < 1 $ and $ B(x) $ be a ball centered at $ x $ of 
the radius $ min(1, \frac{1}{\vert x \vert}) $, then the set $ E $ in $ R^d $
is said to be $ \varepsilon $-thin if $ \forall \ x \in R^d$, we have $\ \frac{\vert E \cap B(x) 
\vert} {\vert B(x) \vert} < \varepsilon $. The strongly annihilating pairs can be
expressed in terms of a decomposition of the spacetime, relating the level set
of the functions $ \lbrace \vert x_i \vert^{a_i} \rbrace_{i=1}^d $.
The above definition of the annihilating pairs is linked with the Heisenberg 
type inequalities and they show the existence of spectral gaps for the 
underlying operators. Furthermore, Hardy type uncertainty principle asserts
that the function $ f $ satisfies the following two inequalities (i) $ \vert
f(x) \vert \leq C (1+ \vert x \vert )^N e^{- \pi \vert x \vert^2 }
$, and (ii) $ \vert \widehat{f}(k) \vert \leq C (1+ \vert k \vert
)^N e^{- \pi \vert k \vert^2 } $. Thence, there exists a polynomial $
P(x) $ of the degree at most $ N $ such that the function $f$
satisfies the property that $ f(x)= P(x) e^{-\pi \vert x \vert^2 } $.

From the perspective of the generalized uncertainty principle, it is worth
considering the problem of the quadratic forms. Let $ q, q^{\prime} \in R^d $ 
be any two nondegenerate quadratic forms, then the distributions $ f \in
S^{\prime}(R^d) $ with $ e^{± \pi q} f \in S^{\prime}(R^d) $ and
$ e^{\pm \pi q^{\prime} }f \in S^{\prime}(R^d) $ describe the space
of the distributions. For a class of $ q, q^{\prime} $, it is possible
to find the conditions that the underlying space of the above distributions
either reduce to $0$ or to a singular distribution. In the case,
when the one of the quadratic form is positive, or when one of them
has the $(d- 1, 1)$ signature, the notion of the uncertainty principle
can be described as a corollary of our general examination of the function 
$ \tilde{\delta}(\Delta x, \Delta k, \epsilon) $. To do so, let us
illustrate the case of the two dimensional Euclidean spaces \cite{BDemange}.
Consider a function $ f $ on $ R^2 $ satisfying (i) 
$ \vert f(x,y) \vert \leq C e^{ −2 \pi a \vert xy \vert } $, and
(ii) $ \vert \widehat{f}(x,y) \vert \leq C e^{−2 \pi b \vert k_1 k_2 \vert } $, 
where $ ab > 1 $, then Ref. \cite{BDemange} states that $ f = 0 $.
A priori, it is not known whether the two functions are in $ L^1
$ or $ L^2 $, thus we need to take the Fourier transform in the
sense of a distribution, instead of the functions. The classical
complex analysis is one of the main tool, which naturally appears
into the picture, once we transform the conditions to the integral
transform. For instance, the Bargmann's representation theory is 
an interesting example. Thus, for a given pair $ (q, q^{\prime}) $, 
the two sets $\lbrace x \in R^d $; $\vert q(x) \vert < A \rbrace $ 
and $ \lbrace k \in R^d ; \vert q^{\prime} (k) \vert < A \rbrace $ 
form a weakly annihilating pair in the sense of the distribution. 
The radar ambiguity functions generalize a class of such results 
and they lead to both the strongly and weakly annihilating sets as
per the following consideration.
Let $ u, v $ be the two $ L^2(R^d) $ functions, then the radar 
ambiguity function associated to the pair $(u, v) $ is defined
by $ A(u,v)(x,y)= \int_{R^d} u(k +\frac{x}{2}) \overline{v(k-
\frac{x}{2}}) e^{2i \pi <k,y>} dk $,\ $ \forall \ x, y \in R^d $.
Furthermore, the radar ambiguity function can be generalized by
elementary changes to the Wigner transform, windowed Fourier
transforms and phase retrival problems. From the viewpoint of 
the distribution theory, we anticipate that the general setting
of the function $ \tilde{\delta} $ yield a class of new distribution
maps for the above duality transformations. However, it is worth mentioning that
a specific determination of the corresponding duality map is expected to involve
the full knowledge of nonperturbative behavior of quantum strings 
or alternative models of the quantum gravity itself.
We leave these issues open for a future examination.

 
From the perspective of the geometric distribution theory \cite{E.B.Davies},
the spectral properties of the Laplacian on a manifold, in comparison
with the periodic Schr\"odinger operators, can be investigated by
defining a periodic manifold $ \mathcal N $ or a Riemannian covering
space of $ \mathcal M $ satisfying the properties of a covering transformation
group $ \Gamma $. A priori, let $ \mathcal N $ be a noncompact Riemannian 
manifold of arbitrary finite dimension $ d \geq 2 $. Then, 
(i) $\mathcal N $ is called a periodic manifold, if the action 
of $ \Gamma $ of the isometries of $ \mathcal N $ on $ \mathcal N $ 
are such that the quotient $ \mathcal M:= \frac{\mathcal N}{\Gamma} $
is a $d$-dimensional compact Riemannian manifold. (ii) A fundamental
domain $ \mathcal D_{\gamma} $ is fixed by an open set $ \mathcal
D \subset \mathcal N $ such that $ \gamma \mathcal D \bigcap
\gamma^{\prime} \mathcal D= \phi $ for all $ \gamma \neq \gamma^{\prime} $ 
with $ \bigcup_{\gamma \in \Gamma}\gamma \mathcal D =
\mathcal N $. What follows next that this setting allows us to
geometrically analyze the periodic operators on a given manifold. 
Let $ (g_{ij}) $ be the metric tensor with the inverse metric tensor
$ (g^{ij}) $, then the pointwise norm of the $1$-form $ d \xi $ in 
the coordinate representation is given by
$ \vert d \xi \vert^2= \sum_{i,j} g^{ij} \partial_i \xi \partial_j \xi $. 
Thus, let us consider an elliptic operator with Laplacian $ \Delta_{
\mathcal N } $ on $ \mathcal N $ acting on a dense subspace of 
Hilbert space $ L^2(\mathcal N) $ with the norm $ \Vert .
\Vert_{\mathcal N} $. Then, Ref. \cite{TKato} shows that the
positive selfadjoint operator $ \Delta_{ \mathcal N } $ can be
defined in terms of a suitable quadratic form $ q_{ \mathcal N } $
by the following expression 
$ q_{ \mathcal N} (\xi):= \Vert d \xi \Vert^2_{ \mathcal N }= \int_{
\mathcal N } \vert d \xi \vert^2 d \mathcal N $, where $ \xi \in
C_c^{\infty} $. Because of the $ \Gamma $-invariance of the
metric on $ \mathcal N $, the Laplacian $ \Delta_{ \mathcal N } $
commutes with the translations on $ \mathcal N $, \textit{i.e.}, $
(T_\gamma \xi)(x):= \xi(\gamma^{−1} x), \ \forall \xi \in
L^2(\mathcal N), \ \forall \gamma \in \Gamma $. In fact, the usual
operator $ \Delta_{ \mathcal N } $ is related with the quadratic
form $ q_{ \mathcal N } $ by the formula $ \langle \Delta_{
\mathcal N} \xi, \xi \rangle= q_{ \mathcal N}(\xi), \xi \in
C_c(\mathcal N) $. Hence, the closure of $ q_{ \mathcal N } $ can
be extended onto the Sobolev space $ H^1(\mathcal N):= \lbrace \xi
\in L^2(\mathcal N) \ \vert \ q_{ \mathcal N}(\xi) < \infty \rbrace $.
Thence, one can find an order relation among the eigenvalues of 
either the Dirichlet Laplacian $ \Delta_{\mathcal D}^{+} $ or the
Neumann Laplacian $ \Delta_{\mathcal D}^{-} $ satisfying the
following maximum/ minimum principle $ \lambda_k^{\pm}= inf_{\lbrace L_k \rbrace}
Sup_{\lbrace \xi \ \in \ \frac{L_k}{\lbrace 0 \rbrace}\rbrace}\frac{q_{\mathcal D}^{\pm}(\xi)} {\Vert u \Vert^2}
$, where the infimum is taken over all possible $k$-dimensional subspaces $
L_ k $ of the corresponding quadratic forms over the domains $ dom(q_{
\mathcal D}^{\pm}) $ with $ dom(q_{ \mathcal D}^{+}) \subset
dom(q_{\mathcal D}^{-}) $. Thus, for purely discrete spectrum 
leveled by the index $ k \in N $, the Dirichlet eigenvalues $
\lambda_k^{+} $ are in general greater than the corresponding
Neumann eigenvalues $ \lambda_k^{-} $.


On the other hand, Mustard has developed a new family of measures 
which are (i) invariant under the group of fractional Fourier transform
and (ii) obey certain uncertainty principle \cite{DavidMustard}. In this
case, the Heisenberg uncertainty principle asserts an inequality
as the measure of the joint uncertainty associated with the function
and its Fourier transform. Mathematically, an uncertainty
principle asserts a reciprocality relation between the spread
of the function $ f $ and the spread of its Fourier transform $
\widehat{f} $. At the infinity, unless both the $ f $ and $ \widehat{f} $ 
decrease to zero (faster than $ \vert x \vert^{\frac{3}{2}} $),
either the $ \Delta (f) $ or the $ \Delta (\widehat{f}) $ becomes
infinite. In such cases, the Heisenberg uncertainty principle 
turns out to be uninformative about the minimum of the other. 
To get rid of the limitations of the Heisenberg uncertainty principle,
we have offered various generalizations to the other physical situations
as well, than the standard quantum systems and possible different measures
of the spread in the $ \Delta (f) $ and $ \Delta (\widehat{f}) $.
One of the such situation can be illustrated in the signal analysis.
Considering the asymptotic dimension of a class of functions which 
are almost band limited and duration limited for which we have
$ \Vert f- \chi_T f \Vert^2 < \epsilon $ and 
$ \Vert \widehat{f}- \chi_{\Omega} \widehat{f} \Vert^2 < \epsilon $,
as $\Omega T \rightarrow \infty $. Thence, from Ref. \cite{DavidMustard} 
one finds the following fractional Fourier transform $ \Im_{\theta}: 
Span\{f\} \rightarrow Span\{\widehat{f}\} $. This provides an 
invariant definition of a class of functions which are considerably
interesting and appropriate in the uncertainty analysis. For the case
of $ R^2 $, Mustard conjectured \cite{DavidMustard} that 
$ S(A,\epsilon)= \lbrace f \ \vert \ min_{\lbrace E \rbrace} \Vert \mathcal W_f-
\chi_{E(A)} \mathcal W_f\Vert^2_{R^2} < \epsilon \rbrace $, where
$ \mathcal W_f $ is the Wigner distribution of $ f $ and $
\chi_{E(A)} $ is the characteristic function of $ A $.
It is the measure theory of the $ f $ and $ \widehat{f} $, given
respectively as $ \Delta (f) $ and $ \Delta (\widehat{f}) $ satisfying
our completely generalized uncertainty principle
$ \Delta (\widehat{f})= \frac{a}{\Delta (f)} + \tilde{\delta}$ is
important from the perspective of the relationship between the function 
$ \tilde{\delta} $ and $ \mathcal W_f $. Thus, the physical perspective 
of the measure theory offers interesting issues towards the existence 
of the minimal and maximal length scales in nature, black hole physics 
and short distance geometries, \textit{viz.} noncommutative geometry, 
noncommutative Clifford manifolds, and discretization of the spacetime. 
For further discussions on variants of the Heisenberg inequality, 
local uncertainty relations, logarithmic uncertainty inequalities,
results pertaining to the Wigner distributions, qualitative 
uncertainty principles, theorems on approximate concentrations,
phase space decompositions, and other functional properties, 
we refer the survey article \cite{GeraldBFollandandAlladiSitaram}.


\subsection{Discretization Of The Spacetime}

The complete generalization of the uncertainty principle involves
the incorporation of higher integer expectation values, \textit{viz.}
$\{ <x^n>^m\ | \ n, m \in Z \} $. Thus, the other existing generalized 
uncertainty principles are only an approximation to our completely
generalized uncertainty principle. For all $ n, m \in Z $,
our generalized uncertainty expression involves the function 
$ \tilde{\delta}(\Delta x, \Delta k, \epsilon) $, which contains
all possible corrections operating in nature. For example, when the
underlying space is compactified on the circle of radius $ r $, then
the Taylor series or the corresponding Laurant expansion would remain
valid, when $ r $ is larger than the dispersion in the state $ \Delta x^2 $.
Thus, by interchanging the roles of the configuration space and the 
corresponding momentum space, one obtains a discrete spectrum in the
position space. To be specific, let $r_p $ be the compactification radius
of the momentum space, then by interchanging the roles of the position and the momentum, 
one finds that there exists a discrete spectrum of the position operator
whose consecutive eigenvalues are separated by $ \frac{1}{r_p} $, and thus
the position space is discretized. Furthermore, if we want to discretize
the configuration space on a chosen scale, say for instance $ l_P $, then 
the radius $ r_p $ is determined in terms of $ l_P $. In this sense,
the parameter $ \tilde{\alpha}_p $ characterizes inequivalent discretizations
of the position operator and thus it provides a finite separation between
its eigenvalues. For a given compactification radius $r_p $, the separation length 
scale is given by $\frac{1}{r_p}$, where we have set $\hbar= 1$. Moreover, 
$ \forall \ d \geq 2 $, the $R^d $ Schr\"odinger operator $ H:= -\Delta + V $ 
possesses a non-trivial set of gaps in its energy spectrum, where  $ V $ is 
a suitable periodic potential. This condition is ensured by the facts that 
(a) $ V $ is periodic $ \Rightarrow \exists $ a basis $ \lbrace \eta_i \rbrace_{i=1}^d $ 
of $ R^d $ such that $ V(x + \eta_i)= V(x),\ \forall i= 1,\ldots ,d $ and (b) let
$ \mathcal D \subset N $ be a fundamental domain, for example usual
parallelepiped with $ \mathcal D= (0, 1)\eta_1 + \ldots + (0,
1)\eta_d $, then the potential $ V $ has a high barrier near the
boundary of the domain $ \mathcal D $, \textit{i.e.}, $ V $ decouples 
as a fundamental domain $ \mathcal D $ from the other neighboring domains 
$ \lbrace \eta_i+ \mathcal D \rbrace_{i=1}^d $\cite{KwangCShin}.
In this sense, for a given periodic potential $ V $, there exists an 
action of $ Z^d $ on $ R^d $ such that $V$ can be completely 
specified on arbitrary fundamental domain $ \mathcal D \subset N $.

The examples of the states, which do not have an analogue with the 
standard Heisenberg uncertainty principle, are those states
which are localized at the discrete eigenvalues of the corresponding
position operator. In the limit where the discrete eigenvalues of the 
position operator are very finely spaced, our analysis of the generalized 
uncertainty principle leads to an interesting relationship between the 
dispersion of $ x $ and $ p $. 
From the above toy model example of a discretized space and 
quantum mechanical consideration, we find that the generalized uncertainty principle
involves the underlying compactification radius. The corresponding momentum 
dependence could be useful in the models exploring the UV limit.
Thus, our approach contains all realistic models of the discrete 
spacetime and the models pertaining to the quantum theory of gravity.
In the limit of stringy uncertainty relation, our analysis
yields a finite minimum length for the fundamental theories.
Notice further that our generalized uncertainty principle contains
not only an infinite series of higher powers of the momentum dispersion, 
but also involves all typical contributions arising from the higher 
order quantities, \textit{viz.} the momentum expectation values, 
such as $\{ <p^n>^m \ | \ n,m \in Z \}$. 
In a simplified situation, the discretization of the space
is implied by a finite compactification momentum.
In a natural way, this enables one to obtain the leading order 
terms of the generalized uncertainty principle. In this limit, a 
detailed incorporation of the gravity need not be necessary in
order to obtain the limiting generalized uncertainty principle. 
In this concern, we have offered an improved understanding towards the 
origin of the generalized uncertainty principles in the quantum gravity theories. 
This follows from the fact that the complete set of real and integer 
valued properties of $ \tilde{\delta}(\Delta x, \Delta k, \epsilon) $
open new scopes for a further discussion. This is in accordance with the 
developments achieved in the string theory \cite{GreenSchwarzWitten},\cite{Polchinski}. 
Namely, the string theory exists, since it is mathematically consistent, 
and its devil exists, since its nonperturbative consistency 
cannot be explicitly proven. In this sense, we believe that the
investigation of the quantum gravity models of the above sorts
could be useful in order to seek the right direction towards the final 
goal to (i) develop the appropriate mathematical frameworks and 
(ii) construct the real basis towards the ultimate theory of everything.


\section{Conclusion}

In this paper, we have analyzed the complete set of extensions and
modifications of Heisenberg uncertainty principle. Thes lies 
in the heart of complex function theory. In particular, following
the framework of complex analysis, we have shown that our theorem
offers a resolution criterion towards the UV/IR mixing problem in gauge
theories. The theorem of the present interest has been obtained by 
proposing the existence of a function $ \tilde{\delta} $. Subsequently,
we have shown that $ \tilde{\delta} $ includes all possible quantum gravity
scaling effects. The above analysis leads to a new stringy uncertainty principle,
accompanied with the fact that the sizes of the underlying strings are bounded by 
the Planck length scale and the size of the universe. Further, in the factorization 
limit of $ \tilde{\delta} $, the complex function $ \tilde{\tilde{\delta}} $ incorporates 
a class of limiting stringy uncertainty principles and offers interesting clues towards 
the understanding of higher order corrections, nonperturbative stringy corrections, 
physics of quantum gravity, black hole physics, the existence of minimal and maximal
length scales, short distance geometry versus string theory, 
Fourier transformations and the theory of distribution. In this sense, 
we have given an attempt towards the completely generalized uncertainty
principle in the light of discrete spacetime, short distance geometries,
\textit{viz.} noncommutative geometry, noncommutative Clifford manifolds
and noncommutative spacetime properties. At the Plank scale, we find that
the function $ \tilde{\delta} $ reveals geometric origin of the fundamental 
$M$-theory and it provides the underlying explanation towards the geometric
origin of the physical $(3+1)$ spacetime. Based on the lattice structure of 
a given spacetime, we anticipate that the underlying principles behind the 
fundamental string theories are in the agreement with the extended scaling 
properties of the function $ \tilde{\delta} $, as proposed in the previous
section. Thus, we have provided an effective framework to incorporate all 
possible dynamics on the same footing.

In the sequel, we have shown that the above generalized uncertainty principle,
which is based on the analysis of $ \tilde{\delta} $, finds strong supports
from the behavior of string theory, noncommutative geometry and loop quantum gravity.
Moreover, there are many implications originating from our generalized uncertainty
principle for both the low and high energy physics. For example, the generalized uncertainty
principle offers modified viewpoints on the statistical mechanics, \textit{e.g.},
the volume of the fundamental cell in phasespace can be determined in a momentum
dependent manner, as per the consideration of the function $ \tilde{\delta} $. Furthermore,
we provided a resolution for the UV/IR mixing problem in quantum field theories
and gauge theories. In an accordance with our result, the quantum gravity 
features are appeared to have a set of novel implications for the statistical 
properties of underlying thermodynamical and quantum statistical configurations.
Thus, the above generalized uncertainty principle renders to be a common feature
of all promising candidates of short distance geometries and quantum theory of gravity. 
From the perspective of string theory, loop quantum gravity and noncommutative
geometry, our analysis gives a deeper insight on the short distance nature
of spacetime. For example, all the above theories call modifications
of the standard Heisenberg uncertainty principle near the Planck scale. 
In this sense, we have demonstrated in general that a modification of the
corresponding dispersion relation is unavoidable in the quantum gravity
scenarios. This indicates a positive sign for the existence of black hole
remnants and thereby induced quantum gravity corrections to the standard
Bekenstein-Hawking entropy and the Hawking temperature of black hole configurations.


From the perspective of the quantum gravity, our generalized uncertainty 
principle analysis generically applies in all possible limiting cases,
\textit{e.g.}, when the momenta are of the order of Planck scale and
the corresponding gravitational effects follows from the limits of 
$ \tilde{\delta} $. In the context of quantum mechanics, it is worth mentioning that
our generalized uncertainty principle describes a set of general consequences for the
quantum gravity, \textit{e.g.}, the first Taylor coefficient of the function
$ \tilde{\delta} $ offers information to determine the minimal length of the theory.
Physically, the above effect can be realized that, if enough mass-energy
is confined to a small region of the space, then a black hole must be formed
\cite{GtHooft}, and thus there must exist the minimal measurable length in nature.
In this concern, if one increases the energy of the colliding particles
beyond the Planck energy, then our analysis anticipates that the short
distance effects are hidden behind the event horizon of a black hole. 
As the energy is increased, the size of the event horizon of the black hole increases 
and thereby an effective minimal length scale arises through the discretization 
of the spacetime \cite{KwangCShin}. From the viewpoint of the UV completion, 
our analysis opens up the following question towards the proper resolution
of the minimum length scale problem. Does the generic appearance of the
minimal length scale in low energy effective quantum gravity theories
survives in the full theory of the quantum gravity?


Along with the fact that the most of the modifications of 
uncertainty relations are either motivated by the general property
of quantum gravity or they analyze the corresponding phenomenological
implications of a limiting generalized uncertainty principle. In this
research, we have formulated the most general uncertainty principle,
which is understood from the holomorphic and anti-holomorphic
properties of the function $ \tilde{ \tilde{\delta}} $. In the sense of 
complex analysis, the function $ \tilde{\delta} $ is completed,
when all the relevant terms of the Taylor or Lorentz expansion of $
\tilde{\tilde{\delta}} $ are included into the consideration. In the factorization limit,
the Lorentz coefficients $\{ \tilde{\delta}_n\ | \ n \in Z^{+}\}$ contain perturbative
information, \textit{viz.} in the sense of $z^{n}$ expansion for a given $z \in C$, 
and the remaining terms $\{ \tilde{\delta}_n\ | \ n \in Z^{-}\}$ provide 
non-perturbative properties, \textit{viz.} in the sense of $z^{-n}$ 
expansion for a given $z \in C$, of the quantum theory.
For a given set of discrete eigenvalues of the position operator,
the above notion follows from the fact that the difference between
the eigenvalues of the position operator gives rise to the minimum length 
scale in the chosen theory. Thus, our generalized uncertainty principle,
obtained from the analysis of the function $ \tilde{\delta} $, 
fits with the notion of the completely discretized spacetime, existence
of the minimum and maximal length scales, quantum geometries and black
hole physics. In the above type of quantum gravity theories,
the particular issue of a further interest could follow from the question:
how to quantize a general classical systems with a given phase space? 
In such cases, we wish to finally mention that the tools of the geometric
quantization \cite{wittengq}, \textit{e.g.}, deformation theory, may open 
new research avenues for further investigations in the theory of gravity,
short distance geometries and quantum physics.

\section*{Acknowledgements}

The author would like to thank Bikash Bhattacharjya, Vinod Chandra, Ashok Garai,
Pratyoosh Kumar, Ravindra Kumar, Pankaj Kumar Mishra, J. Prakash, V. Ravishankar, 
Ashoke Sen, Gautam Sengupta, Ravinder Singh, K. P. Yogendran for discussion and 
comments; and Mohd. Ashraf Bhat, Ashish Kumar Mishra, Nirbhay Kumar Mishra for 
reading the manuscript and making several helpful suggestions.
This work was supported by CSIR, India, under the research grant: 
CSIR-SRF-9/92(343)/2004-EMR-I.

\end{document}